\documentclass{ws-ijmpa}
\newcommand{\bee}{\begin{equation}}
\newcommand{\ee}{\end{equation}}
\newcommand{\beea}{\begin{eqnarray}}
\newcommand{\eea}{\end{eqnarray}}

\newcommand{\etal}{{\em et~al.}}

\newcommand{\lsim}{\mathrel{\lower4pt\hbox{$\sim$}}
\hskip-12.5pt\raise1.6pt\hbox{$<$}\;}

\def\Tr{{\rm Tr}}
\def\fB{f_{\rm B}}

\def\fBs{f_{\rm B_s}}

\def\BBhat{\widehat{B}_{\rm B}}

\def\BBshat{\widehat{B}_{\rm B_s}}

\begin{document}

\markboth{DeGrand}{Lattice QCD}

\catchline{}{}{}{}{}

\title{Lattice QCD at the End of  2003}

\author{Thomas DeGrand}
\address{
Department of Physics,
University of Colorado,
        Boulder, CO 80309 USA
}

\maketitle


\begin{abstract}
I review recent developments in lattice QCD. I first give an overview of its formalism,
and then discuss lattice discretizations of fermions.
We then turn to a description of the quenched approximation and why it is
disappearing as a vehicle for QCD phenomenology.
I describe recent claims for progress in simulations which include dynamical
fermions and the interesting theoretical problems they raise.
I conclude with brief descriptions of the calculations of matrix elements in heavy
flavor systems and for kaons.
\end{abstract}



\section{Introduction}
Lattice regulated quantum field theory supplemented by
numerical simulations is a major source of information about Quantum Chromodynamics
(QCD), the theory of the strong interactions. This article is an
 overview of lattice QCD, with a target audience of 
particle physicists who do not do lattice calculations themselves, but
might be ``users'' of them.
Since lattice QCD is  a mature field \cite{STANDARD},
 most work in lattice QCD represents evolutionary progress, not revolutionary
breakthroughs. But  lattice
QCD is presently going through a transition  phase, as a long-used
approximation to QCD (the quenched approximation) becomes increasingly problematic,
and numerical studies of QCD which include the effects of dynamical quarks
are beginning to produce interesting numbers -- and their own interesting
questions of principle.

Most lattice reviews --like most lattice calculations -- are fixated on
numbers, the values of reduced matrix elements computed at physical values
of quark masses.
Numbers are important, and the lattice can be the best source
of some specific (model-independent) predictions of QCD.
 However, numbers evolve and become obsolete.
I think it will be more interesting to take a more impressionistic view of lattice
QCD, and concentrate on how the calculations are done.
There are a lot of ways that they can fail, and
phenomenologists who are going to use lattice determinations of reduced matrix
elements in their own work ought to be aware of the potential weaknesses in them.

I will begin with a quick overview of lattice methodology: how to go from a continuum
action to a lattice calculation, and then get back to the continuum.
Next, I will describe methods for putting fermions on the lattice.
A lot has happened here since the last textbooks were written.
I will then describe the quenched approximation (which is about to
disappear as a source of quantitative lattice predictions) and tell you
about recent work with dynamical fermions, which is certainly in a very peculiar state.
I will then give an overview of two kinds of standard model tests,
first heavy flavor physics, and then weak matrix elements of kaons.
I will not discuss QCD thermodynamics, lattice-motivated scenarios of confinement
 or chiral symmetry breaking,
or $N_c \ne 3$.

\section{A superficial overview of lattice methodology}
\subsection{Lattice Variables and Actions}

In order to perform calculations in quantum field theory it is necessary to
 control their ultraviolet divergences.
The lattice is a space-time cutoff which eliminates all degrees of freedom
from distances shorter than the lattice spacing $a$. As with any regulator,
it must be removed after renormalization. Contact with experiment only
exists in the continuum limit, when the lattice spacing is taken to zero.
 Other regularization schemes are tied closely to perturbative
expansions: one calculates a process to some order in a coupling constant;
divergences are removed order by order in perturbation theory.  
In contrast, the lattice is a non-perturbative cutoff. Before a calculation begins,
all wavelengths less than a lattice spacing are removed.

All regulators have a price. 
On the lattice we sacrifice all continuous space-time symmetries
 but preserve all internal symmetries, including local gauge 
invariance. 
 This preservation is important for non-perturbative physics. 
 For example, gauge invariance is a property of the continuum theory 
which is non-perturbative, so maintaining it as we pass to the lattice 
means that all of its consequences (including current conservation and 
renormalizability) will be preserved. Of course, we want to make
predictions for the real world, when all symmetries are present. It can be very expensive
to recover the symmetries we lost by going on the lattice.

It is straightforward to construct the lattice version of a field theory.
 One just replaces the
space-time coordinate $x_\mu$ by a set of integers $n_\mu$ ($x_\mu=an_\mu$,
where $a$ is the lattice spacing). 
Field variables $\phi(x)$ are defined on sites $\phi(x_n) \equiv \phi_n$,
 The action, an integral over
the Lagrangian, is replaced by a sum over sites, and derivatives in the Lagrange
density are replaced by finite differences.
\bee
 S = \int d^4x {\cal L}(\phi(x)) \rightarrow a^4 \sum_n 
{\cal L}(\phi_n) \label{2.1}
\ee
and the generating functional for Euclidean Green's functions is
replaced by an ordinary integral over the lattice fields
\bee
Z = \int (\prod_n d \phi_n ) e^{ S}. \label{2.2} 
\ee
Gauge fields carry a
space-time index $\mu$ in addition to an internal symmetry index $a$
($A_\mu^a(x))$ and are associated with a path in space $x_\mu(s)$: a
particle traversing a contour $s$ in space picks up a phase factor
\bee
\psi \rightarrow P(\exp \ ig \int_s dx_\mu A_\mu) \psi
 \equiv U(s)\psi(x). \label{2.3}
\ee
$P$ is a path-ordering factor analogous to the time-ordering
operator in ordinary quantum mechanics. Under a gauge transformation $g$,
$U(s)$ is rotated at each end:
\bee
U(s) \rightarrow V(x_\mu(s))U(s)V(x_\mu(0)^{-1}). \label{2.4} 
\ee
These considerations led Wilson\cite{KEN} to formulate gauge fields
on a space-time lattice, in terms of a set of
 fundamental variables  which are elements of the gauge  group $G$
(I'll specialize to $SU(N)$ for $N$ colors) living
on the links of a four-dimensional lattice, connecting neighboring sites
 $x$ and $x+a \mu$:
$U_\mu(x)$, 
 \bee
U_\mu(x)= \exp (igaT^aA^a_\mu(x))  \label{2.5}
\ee
($g$ is the coupling,  $A_\mu$ the vector potential, 
and $T^a$ is a group generator).

Under a gauge transformation link variables transform as 
\bee
U_\mu (x) \rightarrow V(x) U_\mu (x) V(x+ \hat \mu)^\dagger  \label{2.10}
\ee
and site variables transform as 
\bee
\psi(x) \rightarrow V(x) \psi(x) \label{2.11}.
\ee
We are usually only interested in gauge invariant observables. These will be either
 matter fields connected by  oriented ``strings" of U's 
\bee
\bar \psi(x_1) U _\mu(x_1)U_\mu(x_1+\hat \mu)\ldots  \psi (x_2)  \label{2.13}
\ee
or closed  oriented loops of U's 
\bee
{\rm Tr} \ldots U _\mu(x)U_\mu(x+\hat \mu)\ldots \rightarrow
{\rm Tr} \ldots U_\mu(x)V^\dagger (x+ \hat \mu)V(x+ \hat \mu)
U_\mu(x+\hat \mu)\ldots  .\label{2.12}
\ee

An action is specified by recalling that the classical Yang-Mills
action involves the curl of $A_\mu$, $F_{\mu\nu}$.
Thus a lattice action ought to involve a product of
$U_\mu$'s around some closed contour. Gauge invariance will
 automatically be satisfied
for actions built of powers of traces of U's around arbitrary closed loops,
with arbitrary coupling constants.
If we assume that the gauge fields are smooth, we can expand the link
variables in a power series in  $gaA_\mu's$. For almost any closed loop, the
leading term in the expansion will be proportional to $F_{\mu\nu}^2$.
All lattice actions are just bare actions
characterized by many bare parameters (coefficients of loops). In the
continuum (scaling) limit all these actions are presumed to lie the same universality class,
which is (presumably) the same universality class as QCD with any regularization
scheme, and there will be cutoff-independent predictions from any lattice
action which are simply predictions of QCD.

 The simplest  contour has a perimeter of four links. The ``plaquette action'' or 
``Wilson action'' (after its inventor) is defined as
\bee
 S={{2} \over {g^2}}\sum_n \sum_{\mu>\nu}{\rm  Re \ Tr \ }
\big( 1 - U_\mu(n)U_\nu(n+\hat\mu)
U^\dagger  _\mu(n+\hat\nu) U^\dagger  _\nu(n) \big).  \label{2.6}
\ee
The bare lattice coupling, whose associated cutoff is $a$, is $g^2$.
The lattice parameter $\beta=2N/g^2$ is often written instead of
$g^2=4\pi\alpha_s$.

In the strong coupling limit, the lattice
regularized version of a gauge theory with any internal symmetry group
  automatically confines\cite{KEN}. The interesting question is to
understand what happens as the bare coupling became weaker at fixed
lattice spacing (or for asymptotically free theories, what happens
as the lattice spacing is taken to zero).
 In the strong coupling limit, chiral symmetry is also
spontaneously broken\cite{CHSB}. While it is still not proved, all the
evidence we have is that if the internal symmetry group is $SU(N)$,
 confinement and chiral symmetry persist as the lattice
spacing is taken away.

\subsection{Numerical Simulations}
Lattice QCD has  survived because it is a framework for doing ``exact''
(direct from a cutoff Lagrangian) calculations of QCD, which is simple enough
in principle that it can be taught to a computer.
In a lattice calculation, like any other calculation in quantum field
theory, we compute an expectation value of any observable $\Gamma$
as an average over a ensemble of field configurations:
\bee
\langle{\Gamma}\rangle 
= {1 \over Z} \int [d\phi] \exp(-S)\Gamma(\phi).
\ee
The average is done by numerical simulation: we construct
 an ensemble of states (collection of field variables),
where the probability of finding a particular configuration in the
ensemble is given by Boltzmann weighting (i.~e. proportional to $\exp(-S)$.
Then the expectation value of any observable $\Gamma$ is given simply by
an average over the ensemble:
\bee
\langle{\Gamma}\rangle \simeq \bar \Gamma
\equiv {1 \over N}\sum_{i=1}^N\Gamma[\phi^{(i)}] .
\label{SAMPLE}
\ee
As the number of measurements $N$ becomes large the quantity $\bar \Gamma$
will become a  Gaussian distribution about a mean value, our desired
 expectation value.

The idea  of essentially all simulation algorithms\cite{METRELAX}
is to construct a new configuration
of field variables from an old one.  One begins with some initial field
configuration and monitors observables while the algorithm steps
along. After some number of steps, the value of observables will appear
to become independent of the starting configuration. At that point the
system is said to be ``in equilibrium'' and Eq. \ref{SAMPLE} can be
used to make measurements.

Dynamical fermions are a  complication for QCD.
 The fermion path integral is not a number and a computer can't
simulate fermions directly.
 For $n_f$ degenerate fermion flavors the generating functional for
Green's functions is
\bee
Z = \int [dU][d\psi][d\bar\psi] \exp(-\beta S_G(U) - \sum_{i=1}^{n_f}
\bar \psi M(U) \psi)
\label{ZFERM}
\ee
One formally integrates out the fermions to give a pure gauge action where the
probability measure includes a nonlocal interaction among the $U$'s:
\bee
Z=\int [dU](\det M(U))^{n_f}\exp(-\beta S(U)) 
\ee
or
\bee
Z = \int [dU] \exp(-\beta S(U)
 + {n_f} {\rm Tr} \ln (M(U))
 ) .  
\ee
Generating configurations of the $U$'s involves computing how the action 
changes when the set of $U$'s are varied. The presence of the determinant makes this
problem very difficult. For a pure gauge theory, changing a variable at one location
only affects the action at sites ``near'' the variable, so the  attempt to
update one link variable  on the lattice involves a computational effort independent
of the lattice volume (said differently, the cost of generating
a new configuration scales with the lattice volume). 
However, the determinant is nonlocal, and so in principle updating
one gauge variable would involve an amount of work proportional 
to the lattice volume.

 Typically, this involves inverting
the fermion matrix $M(U)$ (because the change in
 $\log M$ is $d \log M/dU= M^{-1}dM/dU$).
 This is the
 major computational problem dynamical fermion simulations face.
$M$ has eigenvalues with a very large range--
from $2\pi$ down to $m_q a$. The ``conditioning number'' -- the
ratio of largest to smallest eigenvalue of the matrix -- determines
the convergence rate of iterative methods which invert it.
In the physically interesting limit of
small $m_q$ the conditioning number diverges and the
 matrix is said to become ``ill-conditioned.'' The matrix
becomes difficult (impossible) to invert.  At present it is necessary
to compute at unphysically heavy values of the quark mass and to extrapolate
to $m_q=0$. 
(The standard inversion technique today is one of the variants of
the  conjugate gradient algorithm. )

 The tremendous expense of the determinant
is responsible for one of the standard lattice approximations, the 
``quenched'' approximation. In this approximation the back-reaction of the
fermions on the gauge fields is neglected, by setting
$n_f=0$ in Eq. \ref{ZFERM}.
Valence quarks, or quarks which appear in observables, are kept, but no
sea quarks.

All large scale dynamical fermion simulations today generate
configurations using some variation of the microcanonical ensemble. That is,
they introduce momentum variables $P$ conjugate to the $U$'s and integrate
Hamilton's equations through a simulation time $t$
\bee
\dot U = i P U
\ee
\bee
\dot P = -{{\partial S_{eff}}\over{\partial U}} .
\label{ALLAT}
\ee
The integration is done numerically by introducing a time step $\Delta t$.
The momenta are repeatedly refreshed by bringing them in contact with
 a heat bath and the method is thus called Refreshed or Hybrid Molecular
Dynamics \cite{RMD}.
The reason for the use of these small time step algorithms is that for
any change in any of the $U$'s, $ M^{-1}$ must be recomputed.
When Eqn. \ref{ALLAT} is integrated all of the $U$'s in the lattice
are updated simultaneously, and only one matrix inversion is needed per
change of all the bosonic variables.

For special values of $n_f$ the equations of motion can be derived
from a local Hamiltonian and in that case $\Delta t$ systematics
in the integration can be removed by an extra Metropolis accept/reject
step.  This method is called Hybrid Monte Carlo\cite{HMDHMC}.
There are also many variations on these schemes\cite{ALTHMC}.

Recently, there has been a lot of exploration of alternative, large-step
algorithms for dynamical fermions\cite{OTHERS}. While these look promising, none
has been used as much as the molecular dynamics methods.

\subsection{Spectroscopy Calculations}

\begin{figure}
\centerline{\psfig{file=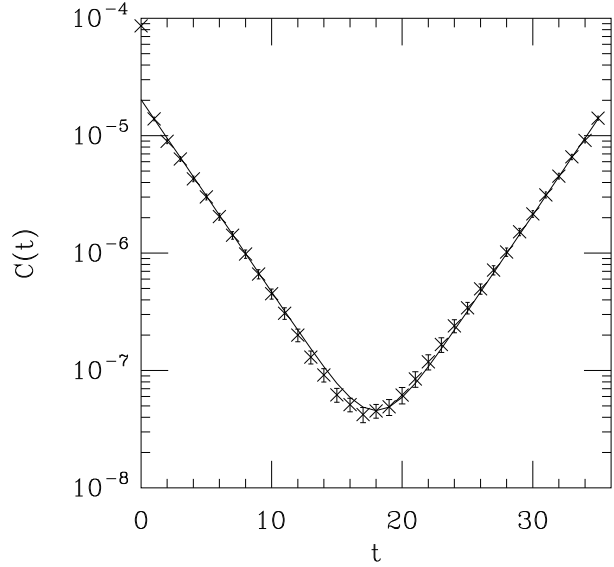,width=8cm}}
\caption{ A ``better than typical'' lattice correlator and its
fit. Periodic boundary conditions convert the exponential decay
into a hyperbolic cosine.}
\label{fig:prop}
\end{figure}

Masses of hadrons are computed in lattice  simulations from the
asymptotic behavior of Euclidean-time
 correlation functions.  A typical (diagonal) correlator can be written as
\bee
K(x) = \langle 0 | O(x) O(0) | 0\rangle  . 
\label{CT} 
\ee
This correlator has its Euclidean-space analog
\bee
\Pi(q) = \int d^4 x \exp(iqx)K(x)
\ee
and if we assume that $\Pi(q)$ is dominated by a sum of
resonances, we expect
\bee
\Pi(q) = \sum_n {{r_n}\over{q^2+m_n^2}}
\ee
where $r_n = |\langle 0|O|n\rangle|^2$.
We can invert the Fourier transform to obtain a prediction for $K(x)$ 
(or some re-weighted analog). For example, lattice people often sum $K(x)$
over a three-dimensional volume of points ($x=\{t,\vec r\}$, sum $\vec r$)
to produce
\beea
C(t) & = & \sum_{\vec r} K(x) \nonumber \\
    & = & \int {{dq_0}\over {2\pi}} e^{iq_0 t} \sum_n  {{r_n}\over{q_0^2+m_n^2}} \nonumber \\
     & = & \sum_n  {{r_n}\over {2m_n}} e^{-m_n t} 
\label{CORRFN}
\eea
At large separation the correlation function is approximately
\bee
C(t) = {{r_1}\over {2m_1}} e^{-m_1 t}
\label{eq:asymp}
\ee
where $m_1$ is the mass of the lightest state which the operator $O$
can create from the vacuum. Fig. \ref{fig:prop} shows an example of a
lattice correlator.
Obviously, the lightest state in a channel is the easiest one to study. Excited
states are hard to work with (their signal goes under the ground state's.)
If the operator has vacuum quantum numbers (an operator coupling to the scalar glueball
is an example), the contribution for the ``real''
lightest state will disappear exponentially into the constant background.
In that case it is also hard to measure a mass.

Most of the observables we are interested in will involve valence fermions.
To compute the mass of a meson
we might take 
\bee
C(t) = \sum_x \langle J(x,t) J(0,0) \rangle 
\ee
where
\bee
J(x,t) = \bar \psi(x,t) \Gamma \psi(x,t) 
\ee
and $\Gamma$ is a Dirac matrix.  The intermediate states $|n \rangle$
which saturate $C(x,t)$ are the hadrons which the current $J$ can create
from the vacuum: the pion, for a pseudoscalar
current, the rho, for a vector current,
and so on. 
Now we write out the correlator in terms of fermion fields 
\bee
C(t) = \sum_x \langle 0 | \bar \psi_i^\alpha(x,t)  \Gamma_{ij}
\psi_j^\alpha(x,t) \bar\psi_k^\beta(0,0) 
\Gamma_{kl}\psi_l^\beta(0,0) | 0 \rangle 
\ee
with a Roman index for spin and a Greek index for color.  We contract creation
and annihilation operators into quark propagators 
\bee
\langle 0 | T(\psi_j^\alpha(x,t) \bar \psi_k^\beta(0,0)) | 0 \rangle
 = G_{jk}^{\alpha \beta}(x,t;0,0) 
\ee
so
\bee
C(t) = \sum_x {\rm Tr} G(x,t;0,0) \Gamma G(0,0;x,t) \Gamma 
\ee
where the trace runs over spin and color indices.
Baryons are constructed similarly. 

In practice, lattice people
do not use such simple operators for interpolating
fields. Instead, they take more complicated operators which model
the wave function of the meson. One way to do this would be
to gauge fix the lattice to some smooth gauge,
like Coulomb gauge, and take $O(\vec x_0,t)= \sum_{x,y}\bar \psi(x,t) \Gamma \psi(y,t)
\Phi(x-x_0,y-x_0)$, where $\Phi$ is our guess for the wave function  (I like Gaussians).
These guesses  often have parameters which can be ``tuned'' to enhance the
signal. It's important to note that the choice of trial
function should not affect the measured mass, it just alters how quickly the leading
term emerges from the sum.

\subsection{Getting rid of the lattice}
The lattice spacing $a$ is unphysical: it was an UV cutoff we introduced by hand to
regularize the theory. The couplings $\{g\}$we typed into the computer
were bare couplings which define the cutoff theory at cutoff scale $a$.
  When we take $a$ to zero we must also specify how the couplings
behave.  The proper continuum limit comes when we take $a$ to zero 
holding physical quantities fixed, not when we take $a$ to zero at fixed lattice couplings
$\{g\}$.
 (It's conventional to
think of $a$ as a function of the bare couplings $\{g\}$, so people actually think about
tuning the $\{g\}$'s to push $a$ to zero. A lattice QCD action will have one marginally relevant
coupling--the lattice analog of the gauge coupling -- plus a set of coefficients
of irrelevant operators. When they
are present, quark masses are relevant operators,
and they must also be tuned as the lattice spacing is varied.
 These quantities characterize the particular lattice discretization.)

On the lattice, if all quark masses are set to zero,
 the only dimensionful parameter is the lattice spacing, 
so all masses scale like $1/a$. Lattice Monte Carlo  predictions are
 of dimensionless ratios of dimensionful quantities.
 One can determine the lattice spacing by
fixing one mass from experiment, and then one can go on to predict any other
dimensionful quantity.
However, at nonzero lattice spacing, all predictions will depend on the lattice spacing.
Imagine computing some masses at several values of the lattice spacing.
(Pick several values of the bare parameters  and
calculate masses for each set of couplings.)
If the lattice spacing is small enough,
 the typical behavior of a ratio will look like
\bee
(a m_1 (a))/(a m_2 (a)) = m_1(0)/m_2(0) + O(m_1a) + O((m_1 a)^2) +\dots \label{SCALING}
\ee
(modulo powers of $\log(m_1a)$).
The leading term does not depend on the value of the UV cutoff.
 That is our cutoff-independent prediction.
Everything else is an artifact of the calculation.
We say that a calculation ``scales'' if the $a-$dependent terms in
Eq. \ref{SCALING} are zero or small enough that one can extrapolate
to $a=0$, and generically refer to all the $a-$dependent terms
as ``scale violations.'' Clearly our engineering goal is to design our lattice
simulation to minimize scale violations.

We can imagine expressing  each dimensionless combination $am(a)$
as some function of the bare coupling(s) $\{g(a)\}$, $am = f(\{g(a)\})$.
 As $a\rightarrow 0$ we must tune the set of couplings $\{g(a)\}$ so
\bee
 \lim_{a \rightarrow 0}
{1 \over a} f(\{g(a)\}) \rightarrow {\rm constant} . \label{2.35}
\ee
From the point of view of the lattice theory,
 we must tune $\{g\}$ so that correlation lengths $1/ma$ diverge.  
This will occur only at the locations of second (or higher) order phase 
transitions.
In QCD  the fixed point is $g_c = 0 $
so we must tune the coupling to vanish as $a$ goes to zero.

One  needs to set the scale by
taking  one experimental number as input.
A complication that you may not have thought of
 is that the theory we simulate on the computer might be
different from the real world, so its spectrum would be different. For example, 
the quenched approximation, or for that matter QCD with two
flavors of degenerate quarks,
 almost certainly does not have the same spectrum as QCD
with six flavors of dynamical quarks with their appropriate masses.
Using one mass to set the scale from one of these
approximations to the real world might not give a prediction
for another quantity which agrees with experiment.

\subsection{Lattice Error Bars}
Numbers presented from lattice simulations come with uncertainties. Phenomenologists
ought to read carefully the parts of the papers which describe the error analysis,
because there are many parts to a lattice number's uncertainty, all different.
Some of the uncertainty is statistical: The sample of lattices is finite.
Typically, the quoted statistical uncertainty includes uncertainty from a fit: it is rare that
a simulation measures
one global quantity which is the desired observable. Usually one has to take
a lattice correlator, assume it has some functional form, and determine the
parameters which characterize the shape of the curve.
The fit function (eq. \ref{eq:asymp} is an example) may be only asymptotic,
and one has to figure out what part of the data is described by one's function.
A complication which enters in at this point is that different quantities measured on the
same set of lattices are typically highly correlated. These correlations have to be taken into
account in the fit.

Dimensionful quantities require (at least) two measurements: an additional 
quantity is needed to set the scale. Again, different quantities may be easier or harder
to measure, so their lattice errors will vary. 
 Clearly, people like to choose quantities which are as
 insensitive as possible  to lattice
artifacts, interpolation in mass, or other analysis issues.
For example, one might want to use a 
hadron mass, or a decay constant, or a parameter associated with the heavy quark
potential, to fix $a$.
Many people use the rho mass, but perhaps this is not a good quantity: in the real
world, the rho is unstable (and broad), so what will happen in a simulations?
The S-P mass splitting in mesons is known to be relatively independent of the quark
mass, but for light hadrons, the P-wave signal is noisy.
Using heavy quark properties to fix the lattice spacing usually requires a different
heavy quark  action,
with a separate simulation. There
are endless arguments in the literature...

This is not the end of the analysis. It may be necessary to extrapolate or interpolate
 one's results to a particular value of the quark mass. One again needs a functional
form. The fit or extrapolation has a statistical uncertainty, but
now systematics begin to creep in: The functional dependence comes from some
 theory. Is the theory
correct? Can it be applied to the whole range of quark masses where the lattice
data exists? 

Additional systematics arise in quenched approximation. Quenched QCD is not
real-world QCD, and their spectra are presumably different.
People publishing quenched numbers sometimes attempt to quote a systematic uncertainty
from the quenched approximation. They might try to do this by using several
different observables to set the lattice spacing, or to fix the quark mass, and
seeing how the final result changes for different observables. The problem
with this analysis is that it assumes that if the parameter values were chosen in the
same way in quenched and full QCD (take the strange quark mass
from the $K^*/\phi$ mass ratio, for example), the desired matrix element would
also be the same in quenched and full QCD. There is no reason for that to be true.
In full QCD there is supposed to be one unique value of the lattice
spacing, so differences from different observables reflect
on the quality of the simulation (assuming that QCD
is in fact the correct description of Nature).

Finally, lattice quantities which are not spectral have scheme dependence. It is
necessary to convert the lattice number to a number in some continuum regularization scheme.
If this scheme matching is done nonperturbatively, the uncertainty 
in doing this is likely to be mostly statistical: the conversion factor comes from its
own simulation. When the matching factor is computed in perturbation theory, most people will 
make ``reasonable'' choices for the way the calculation is organized (picking the scale 
at which the strong interaction coupling constant is determined, for example),
vary their choices in some ``reasonable'' way, and attempt to assign an uncertainty
based on the variation they see.

It is relatively easy to do simulations of gauge theories for any compact internal symmetry
group, or any other bosonic system. Fermionic systems interacting with gauge or matter
fields are feasible without enormous resources for fermion masses down to the strange quark mass, but
become quite expensive below that value. Simulations with massless or nearly massless fermions
remain at the costly frontier. Of course, large $N_c$ simulations scale roughly like
$N_c^3$ from the multiplication of the link matrices. In spectroscopy,  properties
of flavor non-singlet hadrons are easier to compute than those of flavor singlet ones
(disconnected diagrams are noisy), and almost any calculation can be designed to
scale linearly with the number of quark propagators which need to be strung together.

\subsection{Improvement: Why and How}
 Today's 
QCD simulations range from $16^3 \times 32$ to
 $32^3 \times 100$
points and run from hundreds (quenched) to thousands (full QCD) of
hours on the fastest supercomputers in the world. 
 The cost of a Monte Carlo simulation
in a box of physical size $L$ with lattice spacing $a$ and quark mass
$m_q$ scales roughly as
\bee
\big({L \over a}\big)^4 \big({1\over a}\big)^{1-2}\big({1 \over m_q}\big)^{2-3}
\label{COST}
\ee
where the first term in parentheses  just counts the number of sites, the 
second term gives  the cost of
``critical slowing down''--the extent to which successive configurations
are correlated, and the third term gives the cost of inverting the fermion
propagator, plus critical slowing down from the nearly massless
pions. Thus it is worthwhile to think about how to do the discretization,
to maximize the value of the lattice spacing at which useful (scaling) simulations
can be done. 

Remember, the lattice action is just a bare action defined with a cutoff.
No lattice discretization is any better or worse (in principle) than any other.
Any bare action which is in the same universality class as QCD will
produce universal numbers in the scaling limit. However, by clever engineering,
it might be possible to devise actions whose scaling behavior is better,
and which can be used at bigger lattice spacing.

To want to compute the value of some QCD observable via numerical simulation
is to confront a daunting set of technical problems. The lattice volume must be large enough
to contain the hadrons -- and often, to allow them to be located far apart. High statistics
are needed for reasonable accuracy. And the lattice spacing must be small enough
to minimize discretization artifacts. These constraints push simulations
onto large, fast, and often remote supercomputers. This introduces a new
set of problems.
These resources are expensive, so people form collaborations to share them. 
The fastest machines are the newest ones, which are often unstable, or hard to program,
inefficient for all but restricted kinds of computations, 
or all of the above. Finally, because they are so big, lattice projects have a high profile.
They cannot be allowed to fail.

Thus, until the late '90's, most simulations used the simplest discretizations, or
minimal modifications to them. The last few years have seen a slow change in this situation,
basically driven by necessity: standard methods become increasingly dominated
by artifacts
as the light quark masses become smaller and smaller. Two examples illustrate this.

First, groups doing dynamical fermion simulations with staggered fermions
(see Sec. 3)
want to do simulations with two nearly degenerate light quarks (the up and
down quarks) and a heavier strange quark. With this kind of fermion, it is
possible to do simulations with the strange quark mass at its physical
value. However, staggered fermions suffer from flavor symmetry breaking
(see Sec. 3.2), and at today's lattice spacings the standard formulation of
staggered fermions produces a spectrum of mesons in which some of the non-strange
pseudoscalars would be heavier than the strange ones. Then what is a pion, and what is
a kaon?
A more complicated discretization is needed to reduce the flavor
symmetry breaking, so that the non-strange mesons are separated in mass from the
strange ones.

A second example is the discovery of lattice discretizations with flavor symmetry
and improved or exact chiral symmetry (domain wall and overlap fermions). Their
extra computational expense is repaid by their ability to cleanly address problems (mostly
involving chiral symmetry) the standard methods cannot.

The democratization of computing has helped. QCD is not yet possible on 
desk top machines, but a lot of interesting work can be done on clusters for
a cost of a few tens of thousands of dollars.

(Of course, most people will still minimize the amount of evolution of their codes
away from the well-studied algorithms
because the original requirements of large volume and high statistics
have not gone away.)

The simplest organizing principle for ``improvement'' is to
use the  canonical dimensionality of operators as a guide.
Consider the gauge action as an example.
If we perform a naive Taylor expansion of a lattice operator like
the plaquette, we find that it can be written as
\beea
1 - {1 \over 3} {\rm Re \ Tr} U_{plaq} = &
                  r_0 {\rm Tr} F_{\mu\nu}^2  
+a^2 [ r_1 \sum_{\mu\nu}{\rm Tr} D_\mu F_{\mu\nu} D_\mu F_{\mu\nu} +
\nonumber \\
 &  r_2 \sum_{\mu\nu\sigma}{\rm Tr} D_\mu 
F_{\nu\sigma} D_\mu F_{\nu\sigma} + \nonumber \\
 &  r_3 \sum_{\mu\nu\sigma}{\rm Tr} D_\mu F_{\mu\sigma} D_\nu F_{\nu\sigma}]+
 \nonumber \\
 & +O(a^4) 
\eea
The expansion coefficients have a power series expansion in the bare
coupling, $r_j = A_j + g^2 B_j + \dots$.
Other loops have a similar expansion, with different coefficients.
The expectation value of any operator $T$ 
computed using the plaquette action will have an expansion
\bee
\langle T(a) \rangle = \langle T(0) \rangle + O(a) + O(g^2 a) + \dots
\ee
Now the idea is to take the lattice action to be
a minimal subset of loops and systematically remove the
 $a^n$ terms for physical observables order by order in $n$
 by taking the right linear combination of
loops in the action, $S = \sum_j c_j O_j$
with 
$ c_j = c_j^0 + g^2 c_j^1 + \dots$.
This method was developed by Symanzik and 
co-workers\cite{SYMANZIK,WEISZ,LWPURE} in the mid-80's.

Renormalization group ideas have also been used to motivate improvement
 programs\cite{Hasenfratz:1998bb}.

\begin{figure}[h!tb]
\centerline{\psfig{file=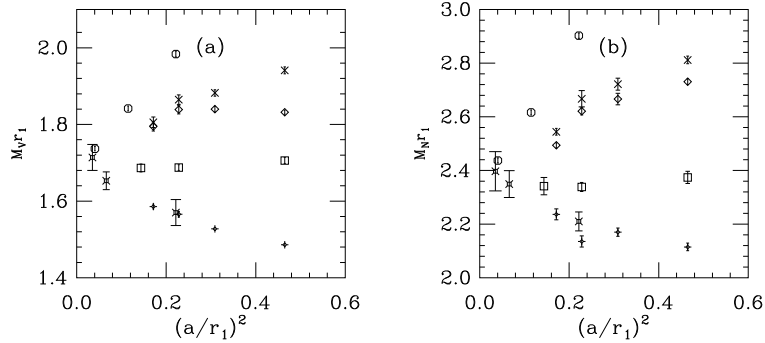,width=10cm}}
\caption{ Lattice calculations of the (a) rho and (b) nucleon mass,
interpolated to the point $m_\pi r_1=0.778$,
as a function of lattice spacing, from Ref. \protect\refcite{MILCSC}. }
\label{fig:scaling}
\end{figure}

If you think about it, all the operators which are being added to
the minimal discretizations are irrelevant operators. So in principle, no
improvement method is really ``wrong,'' although it might happen that
the cost of the improved action is greater than the gain from simulation fidelity.
The final arbiter is a simulation. 
An example\cite{MILCSC}
 of a test of scale violations is shown in Fig. \ref{fig:scaling}.
The x axis is the lattice spacing, in units of a quantity $r_1$,
which is defined through the heavy quark potential:
 $r_1^2 dV(r)/dr|_{r_1}=1.0$, about 0.4 fm. The plotting symbols are for
different kinds of discretizations. The flatter the curve, the smaller
the scale violations. The data labeled by squares shows the best scaling.

A recent innovation for improvement is the use of ``fat links'' as gauge
connections in fermion actions. A standard ``thin link'' action has its
gauge connections built out of a single link variable, 
$S\simeq \bar \psi(x) U_\mu(x) \psi(x+\hat\mu)$. Fat links replace the single
link by a average of paths. For example, a simple blocking (``APE blocking\cite{ref:APEblock}'') is
\beea
V_\mu(x) = &
(1-\alpha)U_\mu(x) \nonumber  \\
& +   \alpha/6 \sum_{\nu \ne \mu}
(U_\nu(x)U_\mu(x+\hat \nu)U_\nu(x+\hat \mu)^\dagger
\nonumber  \\
& +  U_\nu(x- \hat \nu)^\dagger
 U_\mu(x- \hat \nu)U_\nu(x - \hat \nu +\hat \mu) ),
\label{APE}
\eea
with possibly a further projection back to the gauge group.
The parameter $\alpha$ can be tuned.
 There are of course many possibilities
This year,
the most popular choices for fattening include the Asqtad link\cite{OST}
(``$a^2$ TADpole improved'') and the HYP (hypercubic) link\cite{ref:HYP}.

For smooth fields the fat links
have an expansion $V_\mu(x) = 1 + i a B_\mu(x)+\dots$
and the original thin links have an expansion
$U_\mu(x) = 1 + i a A_\mu(x)+\dots$,
where
\bee
\label{eq:linA}
B_\mu(x) = \sum_{y,\nu} h_{\mu\nu}(y) A_\nu(x+y)\ 
\ee
and the convolution function $h_{\mu\nu}(y)$ obviously depends on the fattening.
Smoothing is easiest to see in  momentum space, where
 the connection between the thin and fat link  basically introduces
a form factor\cite{Bernard:1999kc},
$B_\mu(q) = \sum_{\nu} h_{\mu\nu}(q) A_\nu(q)$.

A number of ideas motivate fat links. They all boil down to the simple idea that
a hard lattice cutoff introduces more discretization artifacts than a smoothed cutoff.

Clearly, the trade off is between smoothing the gauge field locally (at the cutoff scale)
 versus erasing physics at long distances. 
Fat links improve perturbation theory, reduce flavor symmetry breaking for staggered
fermions, reduce chiral symmetry breaking for non-chiral discretizations, and
help in the implementation of overlap actions\cite{DeGrand:2002vu}.

\section{Relativistic Fermions on the Lattice}

Finding a lattice discretization for light fermions
 involves yet another problem: doubling.
 Let's illustrate this with free field theory.  The continuum free action is
\bee
S = \int d^4 x [ \bar \psi(x) \gamma_\mu \partial_\mu \psi(x) + m \bar \psi(x)
\psi(x)  ] .\label{2.23}
\ee
One obtains the so-called naive lattice formulation by replacing the
 derivatives
by symmetric differences: we explicitly introduce the lattice spacing $a$
in the denominator and
write
\bee
S_L^{naive} = \sum_{n,\mu} \bar \psi_n \gamma_\mu  \Delta_\mu\psi_n
 + m \sum_n \bar \psi_n \psi_n , \label{2.24}
\ee
where the lattice derivative is
\bee
\Delta_\mu \psi_n = {1 \over {2a}}
(\psi_{n+\mu} - \psi_{n-\mu}).
\ee
The propagator is easy to construct:
\bee
{1\over a} G(p) = (i \gamma_\mu \sin p_\mu  a + ma)^{-1} 
= {{-i \gamma_\mu \sin p_\mu a + ma}\over{\sum_\mu \sin^2 p_\mu a + m^2 a^2}} .
\label{2.25}
\ee
Now the lattice momentum $p_\mu$ ranges from $-\pi/a$ to $\pi/a$.
A continuum fermion with its propagator $(i\gamma_\mu p_\mu + m)^{-1}$
has a large contribution at small $p$ from four modes which are bundled together
into a single Dirac spinor. The lattice propagator has these modes too,
 at $p= (0,0,0,0)$,  but  there are 
other degenerate
ones, at $p = (\pi,0,0, 0)$, $(0,\pi,0,0)$, \dots $(\pi,\pi,\pi,\pi)$. As $a$ goes
to zero, the propagator is dominated by the places where the denominator is small,
 and there are sixteen of these (64 modes in all)
in all the corners of the Brillouin zone.
Thus our action is a model for sixteen light fermions, not one.
This is the famous ``doubling problem.''

The doubling problem is closely connected to the axial anomaly.
Karsten and Smit\cite{Karsten:1980wd}
showed by explicit calculations that the axial charges of the sixteen light fermions
are paired and sum to zero. Many years ago Adler showed that it is not possible
to find a continuum regulator which is gauge invariant for a theory
 with continuous chiral symmetry and
a weak coupling limit with a  perturbative expansion\cite{ADLER},
and the lattice was not immune to this problem.

Nielsen and Ninomiya\cite{NNTHEOREM} codified this constraint in a famous ``no-go'' theorem.
In detail, the theorem assumes
\begin{itemize}
\item{
A quadratic fermion action $\bar \psi(x) H(x-y)\psi(y)$, where $H$ is Hermitian,
has a Fourier transform $H(p)$ defined for all $p$ in the Brillouin zone,
and has a continuous first derivative everywhere in the Brillouin zone.
$H(p)$ should behave as $\gamma_\mu p_\mu$ for small $p_\mu$.
}
\item{
A local conserved charge $Q$ defined as $Q=\sum_x j_0(x)$, where $j_0$ is a function
of the field variables $\psi(y)$ where $y$ is close to $x$.}
\item{$Q$ is quantized.}
\end{itemize}
The statement of the theorem is that, once these conditions hold,
$H(p)$ has an equal number of left handed and right handed fermions
for each eigenvalue of $Q$. 

There is a folkloric version of the theorem  which says that no lattice
action can be undoubled, chiral, and have couplings which extend over
a finite number of lattice spacings (``ultra-locality''). This is actually not what the theorem says.
It is the quantization of the charge which governs whether the theorem is evaded or not.

Ultra-locality is a historical engineering constraint on lattice action design.
What is needed for a proper field theoretic description is locality,
meaning that the range of the action is restricted to be on the order
of the size of the spatial cutoff.
It is believed that having lattice couplings which fall off exponentially
with distance (measured in units of the lattice spacing),
i.e. $S= \sum_{x,r} \bar \psi(x)C(r) \psi(x+r)$ with $C(r) \simeq \exp(-r/\xi)$,
$\xi \propto a$, corresponds
to a local action in the continuum limit, but that slower falloff
(power law, for example) does not. Present day simulations have cutoffs
of 0.2 to 0.05 fm, which is uncomfortably close to physical scales,
and so people who work with non-ultra-local actions worry about 
the range of their actions, and try to tune them to maximize locality.

At the end of their paper, Nielsen and Ninomiya discuss ways to evade the theorem. 
They were not too optimistic about abandoning the quantization of charge, but that
is how overlap and domain wall fermions achieve chirality.

Having said that, it is useful to classify lattice actions into their
folkloric categories: ultra-local actions which are non-chiral and undoubled
(Wilson fermions and their generalizations), ultra-local actions with quantized
chiral charges, which are therefore doubled (staggered fermions), and chiral actions which evade
the Nielsen-Ninomiya theorem, the related cases of domain wall and overlap
fermions.

\subsection{Wilson Fermions (undoubled, non-chiral, ultra-local)}

We can alter
 the dispersion relation so that it has only one low energy solution.  The
other solutions are forced to $E \simeq 1/a$ and become very heavy as $a$
is taken to zero.  The simplest version of this solution, called a
 Wilson fermion, adds  an irrelevant operator, a second-derivative-like term
\bee
S^W = -{r \over {2a}}\sum_{n,\mu}\bar \psi_n(\psi_{n+\mu} -2 \psi_n
+\psi_{n-\mu} ) \simeq a r \bar \psi D^2 \psi \label{2.26}
\ee
to $S^{naive}$.  The parameter $r=1$ is 
almost always used and  is implied when one
speaks of using ``Wilson fermions.'' 
 The propagator is
\bee
{1\over a} G(p) = {{-i \gamma_\mu \sin p_\mu a + m a -r \sum_\mu (\cos p_\mu a -1)} \over
{\sum_\mu \sin^2 p_\mu a + (m  a-r \sum_\mu(\cos p_\mu a -1))^2}} .
\label{2.27}
\ee
It remains large at $p_\mu \simeq (0,0,0,0)$, but the ``doubler modes''
are lifted at any fixed nonzero $r$ so $G(p)^{-1}$ has one four-component minimum.

\begin{figure}
\centerline{\psfig{file=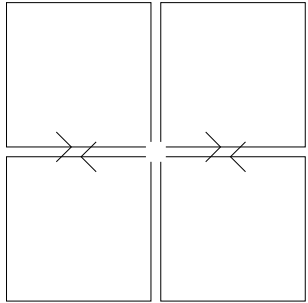,width=4cm}}
\caption{ The ``clover term''.}
\label{fig:clover}
\end{figure}

There are actually two dimension-five operators which can be added to a fermion
action. The Wilson term is just one of them. The other dimension-five
term is a magnetic moment term
\bee
S_{SW} - {{iag}\over 4} \bar \psi (x)\sigma_{\mu\nu}F_{\mu\nu} \psi(x)
\ee
and if both terms are included, their
 coefficients can be tuned so that there are no
$O(a )$ or  $O(a g^2)$ lattice artifacts.
This action is called the``Sheikholeslami-Wohlert\cite{SHWO}''  or
  ``clover'' action because the lattice version
of $F_{\mu\nu}$ (we'll call it $C_{\mu\nu}$ below) is the sum of imaginary parts of
the product of links around the paths shown in Fig. \ref{fig:clover}.
These days, pure Wilson fermions are rarely used, having been replaced by the
 clover action (with various choices of clover term).

With Wilson fermions it is conventional to talk about
``hopping parameter'' $\kappa = {1 \over 2}(m a + 4r)^{-1}$, and to
rescale the fields $\psi \rightarrow \sqrt{2 \kappa} \psi$.  The action for an
interacting theory is conventionally written
\beea
S = & \sum_n \bar \psi_n \psi_n -  \kappa \sum_{n \mu}(\bar \psi_n
(r - \gamma_\mu) U_\mu(n) \psi_{n+ \mu} +
\bar \psi_n(r + \gamma_\mu) U_\mu^\dagger \psi_{n - \mu} ) \nonumber \\
&+ c_{SW} \bar \psi(x) \sigma_{\mu\nu} C_{\mu\nu}\psi(x). \\ \label{2.28}
\eea
From an operational point of view, Wilson-type fermions are closest to the continuum
formulation-- there is a four component spinor on every lattice site for
every color and/or flavor of quark. The simplest bilinears 
($\bar \psi(x)\gamma_\mu \psi(x)$,
for example), could be used as interpolating fields, although in practice more
complicated constructs are used to remove lattice artifacts from simulation
results.

Wilson-type fermions contain explicit chiral-symmetry breaking terms.
This is a source of many bad lattice artifacts. The most obvious is that
 the zero bare
 quark mass limit is not respected by interactions;
the quark mass is additively renormalized.    The value of 
bare quark mass $m_q$  which the pion mass vanishes, is
not known a priori before beginning a simulation. It must be computed.
Simulations with Wilson fermions have to map out $m_\pi$ vs $a m_q$ by direct observation.
 A second serious problem with the loss of
 chiral symmetry breaking
is the mixing of operators which would not mix in the continuum.
This  compromises matrix element calculations.
(See Sec. 6.2 for more discussion.)
Finally, nothing
prevents the Wilson-Dirac operator on a
 gauge configuration
from developing a real eigenmode  $\lambda$ at any value. If $-\lambda$
happened to equal the bare quark mass 
dialed into the program, $D + m$ would be non-invertible. In practice, these
``exceptional configurations'' make it difficult (if not impossible) to push to small values
of the quark mass with non-chiral fermions.
Rare indeed is the Wilson or clover calculation done at pseudoscalar-to-vector
mass ratio much below 0.6.

Besides the simple actions I have described, there are Baroque variations which
are designed to improve either the kinetic properties (dispersion relation),
or chiral properties (minimize additive mass renormalization), or both\cite{FANCY}.

A recent development\cite{Frezzotti:1999vv}
 which removes exceptional configurations is ``twisted
mass QCD.'' This is a scheme for $N_f=2$ flavors in which the lattice
Dirac operator is expanded to be
\bee
D_{twist} = D_W + i \mu \gamma_5 \tau_3.
\ee
The isospin generator $\tau_3$ acts in flavor space. The extra term is called
the ``chirally twisted mass.'' It protects the Dirac operator against 
exceptional configurations for any finite $\mu$: $\det D_{twist}=
  \det(D_W^\dagger D_W + \mu^2)$. It is an amusing exercise to use the
axial transformations $\psi \rightarrow \exp(i\alpha \tau_3 \gamma_5)$,
$\bar \psi \rightarrow \bar \psi \exp(i\alpha \tau_3 \gamma_5)$ to
disentangle what would otherwise appear to be an unfortunate mixing of opposite-parity
operators. I am aware of quenched calculations with this action\cite{TMSIMS},
 but so far, there are no published
 full QCD simulations.

\subsection{Staggered or Kogut-Susskind Fermions (chiral, doubled, ultra-local)}

The sixteen-fold degeneracy  doublers of  naive fermions can be condensed to four
by the local transformation
$ \psi_n \rightarrow \Omega_n \chi_n$,
$ \bar \psi_n \rightarrow \bar \chi_n \Omega_n^\dagger$
where
\bee
\Omega_n= \prod_{\mu=0}^3(\gamma_\mu)^{n_\mu}.
\ee
There are sixteen different $\Omega$'s. Using
\beea
\Omega_n^\dagger\Omega_n &=& 1; \nonumber \\
\Omega_n^\dagger\gamma_\mu\Omega_{n+\hat\mu} &=& (-1)^{n_0 +
   n_1 +\dots +n_{\mu-1}} \equiv \alpha_\mu(n), \nonumber \\
\eea
we rewrite the action as
\bee
S=\sum_n
\bar \psi_n (\gamma_\mu \cdot \Delta_\mu + m) \psi_n = \sum_n \bar \chi_n(\alpha(n)\cdot\Delta + m)\chi_n.
\ee
Written in terms of $\chi$, the action is diagonal in spinor space. 
Although we did the derivation
for free-field theory, it is true for any background gauge field.
$\chi$ is a four-component spinor, but since all components interact independently
and identically, we can reduce the multiplicity of naive fermions by a factor of four simply
by discarding all but one Dirac component of $\chi$. These are ``staggered fermions.''

It is natural to think of the 16 components of a staggered fermion as a fourfold
replication of four Dirac components. The replication is called ``taste.''
``Taste'' is the modern word for what used to be called ``flavor,'' as in
the sentence ``a single staggered fermion corresponds to four flavors/tastes.''

There are (at least) two ways to think about the taste content of
free staggered fermions.
The simplest is just to work in momentum space and
break up the lattice Brillouin zone into 16 components, labeling $s_\mu=0,1$ for each direction,
 with a reduced zone for each component
(the ``central'' one is $ -\pi/(2a) < p_\mu(s) < \pi/(2a$). The massless action is
\bee
S \simeq\sum_s \int_{p_s} \bar \psi(-p)  \sum_\mu i \gamma_\mu \sin(p_\mu) \psi(p)
\label{eq:kspace}
\ee
The ``hypercubic'' decomposition is used in simulations. Break the lattice up into
$2^4$ site hypercubes and bundle the fields in the hypercube together.
Now we define the first and second block derivatives ($b=2a$) by
\begin{equation}
\bigtriangleup_\mu\chi_{n}(N)=\frac{1}{2b}(\chi_n(N+\hat{\mu})
-\chi_n(N-\hat{\mu}))
\end{equation}
\begin{equation}
\Box_\mu\chi_{n}(N)=\frac{\chi_n(N+\hat{\mu})+\chi_n(N-\hat{\mu})-2\chi_n(N)}
{b^2}
\end{equation}
In this basis the action is
\begin{equation}
S \simeq \sum_{x,\mu}b^4\bar{\psi}(x)\left[(\gamma_\mu\otimes I)\bigtriangleup_\mu
+\frac{1}{2}b(\gamma_5\otimes\gamma^*_\mu\gamma_5)\Box_\mu \right]\psi(x)
+mb^4\sum_x\bar{\psi}(x)I\otimes I\psi(x)
\label{stag2}
\end{equation}
The sum over
$x(=Nb)$ runs over all hypercubes of the blocked lattice.
(The notation is (spin $\otimes$ taste).  Taste
 symmetry is four-fold, so the use of Dirac matrices for its generators is natural.)
The presence of the $(\gamma_5\otimes\gamma^*_\mu\gamma_5)$ term shows that the
hypercube decomposition  still has flavor mixing  away from
the $a\rightarrow 0$ limit.

In the continuum, QCD with  four degenerate massless flavors
has  an  $SU(4)_L\otimes SU(4)_R\otimes U(1)_V$
symmetry, which spontaneously breaks to $SU(4)_V$, and 
 the pseudoscalar spectrum consists of 15 Goldstone particles
and a massive (from the anomaly) eta-prime. On the lattice,
taste and spin rotations are replaced by shifts and rotations in the hypercube.
The continuous symmetries of the continuum are broken down to discrete
symmetries. In particular, taste symmetry is broken.
 Only a $U(1)_V \otimes U(1)_A$ survives, and the spontaneous breaking of
the  lattice analog of the flavor non-singlet axial symmetry
produces a single Goldstone boson at nonzero $a$. The other ``would-be'' Goldstones
are non-degenerate pseudoscalar states whose mass goes to zero in
the continuum limit.
(QCD with $N_f$ flavors of staggered fermions has an internal 
$SU(N_f)_L\otimes SU(N_f)_R$ chiral symmetry.)
The $U(1)$ chiral symmetry protects the quark mass from additive renormalization,
and staggered fermions are preferred over Wilson ones in situations in which
the chiral properties of the fermions dominate the dynamics.

There are two complementary ways to think about taste breaking. In momentum space,
taste mixing occurs in perturbation theory via the 
 emission and absorption of gluons with
momentum near $q_\mu = \pi/a$. These kick quarks from one $s_\mu$ momentum sector
into another one. In the hypercube basis, different tastes sit on different locations
in the hypercube. The environment of link variables varies across the hypercube,
so different tastes see different local gauge  fields.

\begin{figure}
\centerline{\psfig{file=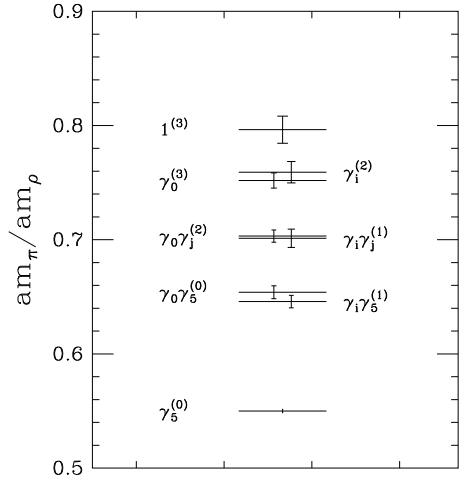,width=6cm}}
\caption{An example of flavor symmetry breaking in an improved
staggered action. The different $\gamma$'s are a code for the various
pseudoscalar states.
Data are from Ref. \protect\refcite{OST}. For an explanation of
the splitting, see Ref. \protect\refcite{SHARPLEE}.}
\label{fig:flav}
\end{figure}

The pseudoscalar spectrum shows an interesting (approximate) degeneracy
(see Fig. \ref{fig:flav} for an example), first described by Sharpe and Lee\cite{SHARPLEE}.
It is reminiscent of the pattern of splittings of the energy levels of
an atom in a crystal lattice.

Today's approach to staggered fermion simulations exploits the conversion of a $U(1)$ chiral symmetry
for a single  staggered flavor (with its single would be Goldstone boson)
 into the $SU(N_f)$ chiral symmetry of $N_f$ flavors (with $N_f^2-1$ Goldstones).
QCD with three flavors of
quarks is treated with a lattice model with three flavors of staggered fermions,
each with its own mass. Each flavor would correspond to four tastes. Three
of the four tastes (per flavor) have to be excised from one's predictions.
How that is done is different for valence and sea quarks, and will be described below.

This decomposition has been around since the mid-90's\cite{Patel:1992vu,Sharpe:1994dc}.
 It was introduced in order
to understand weak matrix elements of staggered fermions.
Recent techniques for staggered fermions, which make heavy use of chiral Lagrangians,
have brought it to greater importance.

Taste symmetry violation for a single staggered flavor basically means that interactions mix
the different tastes. For the connection between staggered flavor and physical flavor to
work,  taste symmetry violation must be minimized.
People attack this problem in two ways.

First, they modify the action to suppress taste violations. This is commonly
done by replacing the thin links of the fermion connection
by a fat link.  The ``Asqtad'' fat link action\cite{OST}  (which also adds a third - nearest -
neighbor coupling to improve the dispersion relation) has seen the most extensive
use. 
Other choices\cite{ref:HYP} reduce taste violations more.

Second, lattice data is analyzed including the effects of
taste violations. The key to doing this was provided by the Sharpe and Lee\cite{SHARPLEE}
analysis of taste mixing,
whose construction of a low energy chiral effective theory including explicit
taste-breaking interactions  predicted the
degeneracies shown in Fig. \ref{fig:flav}. Their work was generalized by Aubin
and Bernard\cite{Aubin:2003uc}
to $N_f$ flavors of staggered fermions ($4N_f$ tastes).
One introduces a field $\Sigma=\exp(i\Phi / f)$, a  $4N_f \times 4N_f$
matrix, and $\Phi$ is given by:
\begin{eqnarray}\label{eq:Phi}
        \Phi = \left( \begin{array}{cccc}
                 U  & \pi^+ & K^+ & \cdots \\*
                 \pi^- & D & K^0  & \cdots \\*
                 K^-  & \bar{K^0}  & S  & \cdots \\*
                \vdots & \vdots & \vdots & \ddots \end{array} \right),
\end{eqnarray}
The entries are the meson fields composed of different staggered flavors, and
 are traces over the 16 taste-product $q\bar q$  bilinears of
each staggered flavor. The mass matrix is
\begin{eqnarray}
        {\mathcal{M}} = \left( \begin{array}{cccc}
                m_u I  & 0 &0  & \cdots \\*
                0  & m_d I & 0  & \cdots \\*
                0  & 0  & m_s I  & \cdots\\*
                \vdots & \vdots & \vdots & \ddots \end{array} \right),
\end{eqnarray}
The Lagrangian is
\begin{eqnarray}\label{eq:final_L}
        {\mathcal{L}} & = & \frac{f^2}{8} {\rm{Tr}}(\partial_{\mu}\Sigma
        \partial_{\mu}\Sigma^{\dagger}) -  
        \frac{1}{4}\mu f^2 {\rm{Tr}}( {\mathcal{M}} \Sigma+{\mathcal{M}}\Sigma^\dagger) \\
        &  & + \frac{2m_0^2}{3}(U_I + D_I + S_I + \cdots)^2 + a^2 {\mathcal{V}},
\end{eqnarray}
where the $m_0^2$ term weighs the  analog of the flavor singlet $\eta'$.
(The ``$I$'' subscripts display that this involves the
taste singlet term for each flavor.)
 The $a^2 {\mathcal{V}}$
term is the taste-breaking interaction, a sum of terms quadratic in  $\Sigma$ with
various taste projectors, parameterized by six coefficients (only one is big).
Now one computes ``any'' desired quantity with this Lagrangian, typically to
one loop, as a function of quark masses and all other coefficients.
Parameters of Nature are determined when mass-dependent Monte Carlo data is fit
to this functional form.  For example, a one-loop fit to $m_{PS}/(m_1+m_2)$ for the
true would-be Goldstone boson made of  quarks of mass $m_1$ and $m_2$
 would involve $\mu$, $f$, two Gasser-Leutwyler parameters, three otherwise
unconstrained lattice parameters, and involves chiral logarithms whose arguments
are all the observed pseudoscalar masses. Fits to $f_\pi$ or $f_K$ are similar.

It is crucial to analyze staggered data this way. In 2001 the MILC collaboration tried and 
and failed to fit their pseudoscalar masses to the continuum chiral logarithm formula.
The same data, with fits which include taste violations, forms part of their
recent high precision calculation of hadronic parameters (see Sec. 5).

\subsection{Evading the no-go theorem -- domain wall and overlap fermions}
Two related schemes allow one to evade the Nielsen-Ninomiya theorem. The first action
is the  ``domain wall fermion\cite{KAPLAN,FURSH}.'' It is
a variation on the idea that  a fermion coupled to a
scalar field which interpolates between two minima
 (a soliton) will develop a zero-energy chiral mode
bound to the center of the soliton. Embed QCD is a five-dimensional  brane world
with a kink. This is a discretized fifth dimension. The gauge fields
remain four dimensional. A chiral fermion will sit on the four dimensional face of
the kink. This is the chiral ``domain wall fermion.'' As Kaplan puts it,
``The extra dimension is the loophole in the Nielsen-Ninomiya theorem through
which the fermions have wriggled.''

Several groups have well-developed programs of QCD phenomenology, both quenched
and dynamical,
with domain wall fermions\cite{BLUM}. We will encounter their results in the
section of this review dealing with kaon physics. 

The practical consideration which must be dealt with in domain wall simulations is
that in the computer, the fifth dimension is not infinite. There will be an
 anti-kink somewhere else in the
fifth dimension  or some equivalent boundary surface, with an opposite-chirality fermion
pinned to it. As long
 as the two kinks are far away, the fermionic chiral mode on the kink doesn't see the
mode on the anti-kink and the
4-d theory on the kink
will remain chiral. But if the anti-kink is too close (typically, if  the  fifth dimension 
is too small or the fermion eigenmodes on the kink are insufficiently localized
in the fifth dimension) the modes mix and chiral symmetry is broken. 
The fermion will be observed
to pick up a small additive mass renormalization.
 How close is
 ``too close'' is (yet) another engineering question, and is dealt with in the usual way (by
modifying lattice discretizations).

The four-dimensional analog of this formulation uses the Ginsparg-Wilson relation
\cite{ref:GW}''
\bee
\gamma_5 D(0) + D(0)\gamma_5 = {1\over{2x_0}} D(0)\gamma_5 D(0).
\label{eq:GW}
\ee
Its explicit realization  is Neuberger's
``overlap fermion\cite{ref:neuberfer}.''
The massless overlap Dirac operator is
\bee D(0) = x_0(1+ {z \over{\sqrt{z^\dagger z}}} )
\label{eq:overlap}
\ee
where $z = d(-x_0)/x_0 =(d-x_0)/x_0$ and $d(m)=d+m$ is a massive ``kernel''
 Dirac operator for mass $m$. $d$ can be (almost) any undoubled lattice Dirac operator.
The chiral symmetry of these remarkable actions 
prevents additive mass renormalization, preserves all continuum Ward identities (up
to contact terms) and knows about the index theorem\cite{ref:HLN}.

The eigenmodes of $D(0)$ sit on a circle of radius $x_0$ 
centered at $(x_0,0)$ in the complex plane. The zero mode is chiral (as is the
mode at $(2x_0,0)$) and the complex modes are non-chiral and paired (with complex
conjugate eigenvalues). The Nielsen-Ninimiya theorem is evaded because the chiral
charges are not quantized.  One way\cite{GSOV} to see this is to realize
that the action $S = \bar\psi D(0)\psi$ is invariant under the gauge field dependent
axial transformation\cite{LUSCHER}
\bee
\delta \psi = T \hat \gamma_5 \psi =T  \gamma_5 (1 -  {{ D}\over x_0})\psi; \  \  \
\delta \bar \psi =  \bar\psi \gamma_5 T
\label{eq:ROT}
\ee
where $T$ is a $U(N_f)$ flavor rotation generator. Directly from
the Ginsparg-Wilson relation, $(\hat \gamma_5)^2=1$, so the combination
of this axial vector transformation, plus the usual vector one, generates a conventional
current algebra. The fermion measure is not invariant under the $U(1)_A$ version of Eq. \ref{eq:ROT},
and this leads to the anomaly.  For small $D$, Eq. \ref{eq:ROT} is the usual chiral rotation.
This is the situation for the low eigenmodes of the Dirac operator. But at the other corners
of the Brillouin zone $D\simeq 2x_0$, $\hat \gamma_5$ flips sign,
 and the transformation does not correspond to a chiral rotation.
One can also consider a symmetric version of Eq. \ref{eq:ROT}:
\bee
\delta \psi =  T  \gamma_5 (1 -  {{D}\over {2x_0}})\psi; \  \  \
\delta \bar \psi =  \bar\psi (1 -  {{D}\over {2x_0}}) T \gamma_5
\label{eq:ROT2}
\ee
At the ``far corners'' of the Brillouin zone, where $D=2x_0$, this transformation
vanishes. Either way, the eigenvalue of the chiral charge is not quantized, it varies with $D$.

 The massive overlap Dirac operator is
conventionally defined to be
\bee
D(m_q) = ({1-{m_q \over{2x_0}}})D(0) + m_q
\ee
and it is also conventional to define the propagator so that the chiral
modes at $\lambda=2x_0$ are projected out,
\bee
\hat D^{-1}(m_q) = {1 \over {1-m_q/(2x_0)}}(D^{-1}(m_q) - {1\over {2x_0}}) .
\label{SUBPROP}
\ee
This also converts local currents into order $a^2$ improved operators\cite{Capitani:1999uz}.

No overlap action is ultra-local\cite{Horvath:1999bk} .
Satisfying the Ginsparg-Wilson relation requires an
action which has connections spread out to an arbitrary number of lattice spacings.
 For sufficiently smooth gauge configurations, Hernandez,
Jansen, and Luscher have
 proved that overlap actions are local\cite{Hernandez:1998et}.
Golterman and Shamir\cite{Golterman:2003qe} have presented a convincing
argument that sufficiently rough gauge configurations can drive overlap and
domain wall fermion actions nonlocal. It is presently an item of debate, how much
today's simulations are corrupted by non-locality.

The hard part of an overlap calculation is the
 ``step function'' ($\epsilon(z)= \gamma_5 z/\sqrt{z^\dagger z}$).
There are various tricks for evaluating it, basically as polynomials in  $z$
(Chebyshev polynomials) or as a ratio of
 polynomials $A(z)/B(z) \simeq \gamma_5 z\sum(1/(z^\dagger z + c_n)$. 
(For discussions of overlap technology, see Refs. \refcite{ref:FSU,vandenEshof:2002ms}.)
The degree of success of these evaluations typically depends on
the conditioning number of $z$, and to
improve that, people remove low eigenmodes of $z$ from the evaluation
and treat them exactly. Overlap calculations typically cost a factor of 50-100 as much as 
ordinary Wilson calculations, but to quote that fact alone is unfair: they can be used 
to study small quark mass (chiral) physics with full flavor symmetry
at quark masses where Wilson-type actions simply fail due to exceptional
configurations.
Unfortunately, overlap fermions are still too expensive for anything but
extremely tentative dynamical fermion simulations.

There is of course an aesthetic (or engineering) debate among lattice practitioners over whether it is
better to do simulations which have approximate chiral symmetry or exact
chiral symmetry. The former are generally computationally less expensive, but
they have to be checked against the appearance of unwanted effects.
With an exact algorithm, one might sleep better at night knowing that the
calculation does not have chiral artifacts, but it might also happen that
one is simply unable to generate an interesting data set for analysis.

Because simulations with domain wall and overlap fermions are so expensive,
there is a tendency to cut corners, either by making the lattice spacing uncomfortably large,
or the simulation volume too small. (The author's calculation of $B_K$ using overlap
fermions is done at
a small enough volume, that baryon masses are clearly affected--they are pushed to
artificially high mass by the squeezing of the box.)

One can turn a bug into a feature by simulating QCD in the ``epsilon regime\cite{EPSILON}.''
This is QCD in a box whose length $L$ is small compared to the pion Compton
wavelength, $m_\pi L << 1$. It is not small simulation volume, since
$L$ should also be large compared to any `typical'' QCD confinement scale.
This year we have begun to see simulations (mostly with overlap fermions)
in this regime\cite{RECENTEPS}. By analyzing the behavior of hadron correlators, and matching
onto chiral Lagrangian calculations, also done in the epsilon
regime, low energy properties of QCD may be extracted.

The descriptions of domain wall and overlap fermions did not make them look particularly similar.
The connection comes when one takes domain wall fermions and integrates out the bulk fields to
construct a four dimensional effective action of the
 light ``pinned'' fields\cite{ref:neuberfer,OUT},
\bee
D = 1 + \gamma_5{{(1+H)^{N_5} - (1-H)^{N_5} } \over {(1+H)^{N_5} + (1-H)^{N_5} } }
\ee
where $H=\gamma_5z/(2+z)$ and $z$ was defined above. As the length of the fifth
dimension becomes infinite and the lattice spacing in the fifth dimension vanishes,
 this Dirac operator becomes the overlap operator.

Finally, different kinds of fermions
have different constraints
on the number of flavors which can be simulated with various algorithms.
Consider a single flavor of staggered (four tastes) or Wilson-type fermions (one flavor), with a
massive Dirac operator $M$. In either case $M^\dagger = \gamma_5 M \gamma_5$.
The functional integral for a theory with a single flavor is just $\det  M$.
The massless staggered fermion Dirac operator
$M(m=0)$ is antihermetian so its eigenvalues lie
along the imaginary axis, and they come in pairs $\pm i \lambda$. For massive
fermions, $\det M$ is a product of $(i\lambda +m)(-i\lambda+m)$ factors, 
thus it is positive-definite for all gauge configurations.
Standard algorithms
 (Refreshed Molecular Dynamics\cite{RMD}, Hybrid Monte Carlo\cite{HMDHMC}) 
actually work with $M^\dagger M$, which would redouble the
number of flavors. But because $M^\dagger M$ only connects ``even'' sites on the
lattice with other ``even'' sites (and `odd'' with ``odd''),  one can take the
square root
of the operator by removing the  fermion fields on
all the ``even'' (or ``odd'') sites.

Wilson fermions have eigenvalues which are real, and complex conjugate pairs.
The determinant from the complex conjugate pairs is basically just like the
staggered determinant; it is a product of $(m+ \lambda_r)^2 + \lambda_i^2$. This part
is positive definite. However, nothing protects the real eigenvalues from taking any value, so the
product of real eigenvalues could have any sign, and could change sign as the gauge field
evolves. With two degenerate flavors (or $N_f=2j$), one can get a positive determinant
because $\det M = \det M^\dagger$ and $(\det M)^2 = \det(M^\dagger M)$.

Furman and Shamir\cite{FURSH} have proved that one can simulate
any number of flavors with domain wall fermions.
For overlap fermions, the real eigenvalues are restricted to be either zero or a positive number,
and so one could also simulate any number of fermions, each one with its own
$\det M$, with no problem with positivity\cite{Bode:1999dd}.

\section{The Quenched Approximation in Health and Sickness}
\begin{figure}[thb]
\centerline{\psfig{file=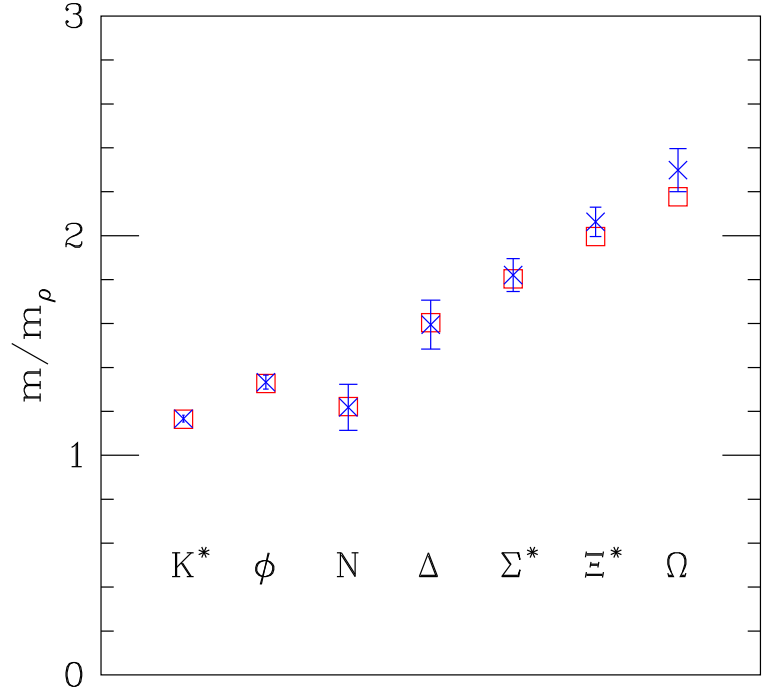,width=10cm}}
\caption{
Comparison of quenched results from Ref. \protect\refcite{Butler:em}
with experiment.}
\label{fig:wein}
\end{figure}

For many years, calculations of spectroscopy and matrix elements in QCD have
 used the ``quenched approximation''
in which the fermion determinant  is completely discarded from the functional integral.
This approximation was adopted pretty much for reasons of expediency, because
numerical simulations in full QCD were simply too expensive. From its
earliest days it has produced results for known quantities which were in 
remarkable agreement with experiment. The results of one very careful
quenched simulation\cite{Butler:em} of the light hadron spectrum are shown in Fig. \ref{fig:wein}.
One could not ask for anything better.

The quenched approximation has many of the ingredients of successful hadron phenomenology.
Quarks are confined (with a linear confining potential if they are heavy).
Chiral symmetry is spontaneously broken.
In it, all states are (at first glance) infinitely narrow,
 because $q \bar q$ pairs cannot pop out of the vacuum.
One might also try to ``justify'' the quenched approximation by an appeal to the quark
model: in the quenched approximation, all mesons are $q \bar q$ pairs, and all baryons
are $qqq$ states.
This also appears to be rather similar to the large-$N_c$ limit of QCD.

However, the situation has changed. For about ten years people have identified
specific situations where the quenched approximation would 
give qualitatively different behavior
than QCD with any nonzero number of sea quarks. This behavior
now needs to be dealt with -- or
 is beginning to be seen -- in simulations.

The best way to see what is going on is to
consider the low energy limit of QCD, and not to think about quarks and gluons, but
in terms of an effective field theory
theory of QCD, described by chiral Lagrangians in which the would-be
Goldstone bosons are fundamental fields. These Lagrangians have
a set of bare parameters (quark masses, $f_\pi$, the quark condensate $\Sigma$, $\dots$).
As far as the chiral Lagrangian is concerned, these are fundamental parameters.
As far as QCD is concerned, one could compute these parameters from first
principles (for example, $f_\pi m_\pi = \langle 0 |\bar \psi \gamma_0\gamma_5
\psi|\pi\rangle$),
this would fix the parameters of the chiral Lagrangian, and then one could throw away the
 lattice and compute low energy physics using the
chiral Lagrangian. Quenched QCD and QCD with nonzero flavor numbers are different theories
and their low energy parameters will be different. But there is more.
In full QCD the eta prime is heavy and can be
decoupled from the interactions of the ordinary Goldstone bosons. In quenched QCD
the eta prime is not really a particle. The would-be eta prime
gives rise to ``hairpin insertions'' which pollute essentially all predictions.

\begin{figure}[thb]
\centerline{\psfig{file=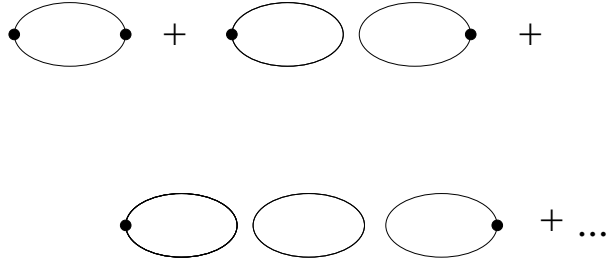,width=8cm}}
\caption{
The eta-prime propagator in terms of a set of annihilation graphs
summing into a geometric series
to shift the eta-prime mass away from the mass of the flavor non-singlet
pseudoscalar mesons. In the quenched approximation, only the first two
terms in the series survive as the ``direct'' and ``hairpin'' graphs.
}
\label{fig:hairpin}
\end{figure}

Let's consider the eta prime channel in
full QCD and quenched QCD in various extreme limits.
In ordinary QCD, the eta prime
propagator includes a series of terms in which the flavor singlet
$q\bar q$ pair annihilates into some quarkless state, then reappears, over and over.
This is shown in Fig. \ref{fig:hairpin}.
 Let's assume
we have $N_v$ valence quarks, $N_s$ sea quarks, and $N_c$ colors.
(The case $N_v \ne N_s$ with $N_s \ne 0$ is called ``partial quenching'' as opposed
to the $N_s=0$ quenched approximation.)
 Let's sum the geometric series for the eta prime propagator
\bee
\eta'(q) = C(q) -H_0(q) + H_1(q) + \dots
\ee
where $C(q) = 1/d$, $d=q^2+m_\pi^2$, is the ``connected'' meson propagator, the
same as for any other Goldstone boson. $H_n$ is
the $n$th order hairpin (with $n$ internal fermion loops). 
 Simple $N_c$ counting appropriate to the large-$N_c$ limit (taking $g^2 N_c$ fixed),
 with flavor and color singlet sources and sinks,
shows that $C(q) \propto N_c^0 N_V^0$ and
 $H_n(q)\propto 1/N_c^n$. 

 The lowest order hairpin is
\bee
H_0(q) = {1\over d} {N_v \over N_c} \lambda^2 {1\over d}
\ee
where $\lambda^2$ is $O(N_c^0)$.
Each extra loop gives another factor of $ -N_s\lambda^2 /(N_c d)$. Summing the
geometric series, we find
\bee
\eta'(q) = {1\over d}(1 - {N_v \over N_s}) + {N_v \over N_s}{1\over{d+\lambda^2N_s/N_c}}.
\label{eq:etaprimepole}
\ee

This formula has a number of interesting limits. First, if $N_v=N_s=N_f \ne 0$,
we see that we have generated a massive $\eta'$ propagator with
 $m_{\eta'}^2=\lambda^2 N_f/N_c$.

The next  case is $N_v < N_s$. There we have both the Goldstone
mode and the eta prime propagator. This is also as expected: think of
the $SU(3)$ flavor symmetry limit and imagine a source $\bar u u + \bar d d$:
it couples to a mixture of the eta (a Goldstone) and the eta prime.

Now we could take $N_c$ to infinity at fixed $N_f$. With 
  $N_c$-scaling of the vertex,
the eta-prime mass falls to zero, and it becomes a ninth 
Goldstone boson as $U(1)_A$ is restored. At any finite $N_c$, the
eta prime
is still an ordinary particle.

However, the quenched limit is different--it is $N_s=0$ first. In that case
(of course)
\bee
\eta'(q) = {1\over d}  - {1\over d} {N_v \over N_c}\lambda^2 {1\over d},
\label{eq:etaq}
\ee
In the eta prime channel there is an ordinary (but flavor singlet) Goldstone
boson and a   new contribution--a double-pole ghost (negative norm)
state. In the $N_c=\infty$ limit, the double pole decouples, but finite $N_c$ quenched QCD
remains different from finite-$N_c$ full QCD. The limits of large $N_c$ and
quenching don't commute.

Similar behavior will occur in any other flavor singlet channel (think
about the $\omega$ meson), of
course. Lattice people haven't talked  about it because
the signals are noisier, and because there is no chiral Lagrangian paradigm.

The double pole would reappear if the sea quarks and valence quarks had different masses.
Its residue would be proportional to the difference of the sea and valence pseudoscalar
squared masses. A convenient form of partial quenching is in fact to compute
hadronic properties for one set of sea quark masses (because each set is expensive)
and many values of valence quark masses (because each set is cheap),

(What is the internal flavor symmetry group of partially quenched QCD? The determinant
of  the valence  quarks is not present in the functional integral. One can cancel
the determinant by introducing $N_v$ valence bosonic quarks, since the power
of the determinant is negative for bosons. The flavor group becomes\cite{Morel,BG}
a graded group
$SU(N_v+N_s | N_v)_L\otimes SU(N_v+N_s | N_v)_R$ which spontaneously
breaks to $SU(N_v+N_s | N_v)$.
The low energy effective Lagrangian has additional mesons, corresponding
to bound states of quarks and bosonic quarks.)

Where the eta-prime comes in  is in the
calculation of corrections to tree-level relations\cite{BG,Sharpe:1992ft}.
 These are typically dominated
by processes with internal Goldstone boson loops,
contributing terms like
\bee
\int d^4 k G(k,m) \simeq ({m \over {4\pi}})^2 \log({{m^2}\over{\Lambda^2}})
\ee
(plus cutoff effects). The eta-prime hairpin can appear in these loops,
replacing $G(k,m) \rightarrow -G(k,m) \lambda^2 G(k,m)$ and altering the chiral logarithm.
Thus, in a typical observable, with a small mass expansion
\bee
Q(m_{PS}) = A(1 + B{ m_{PS}^2 \over f_{PS}^2 }\log m_{PS}^2 )+ \dots
\label{eq:eq001}
\ee
quenched and $N_f=3$ QCD 
can have different coefficients (different $B$'s in Eq. \ref{eq:eq001}), seemingly
randomly different.  (Quenched $f_\pi$ has no chiral logarithm while it does in full QCD,
the coefficients of $O_+$, the
 operator measured for $B_K$, are identical in quenched and full QCD,
etc.)
Even worse, one can find a different functional form. For example, the
relation between pseudoscalar mass and quark mass in full QCD is
\bee
m_{PS}^2 = Am_q(1 + {{m_{PS}^2}\over{8\pi^2f^2}}\log(m^2/\Lambda^2)]+\dots.
\ee
In quenched QCD, the analogous relation is
\bee
(m_{PS})^2/(m_q) = A [1 - \delta (\ln ( m^2/\Lambda^2) +1 )] + \dots
\label{eq:log}
\ee
where $\delta=\lambda^2/(8\pi^2  N_c f_\pi^2)$ is expected to be about 0.2
using the physical $\eta'$ mass.
This means that $m_{PS}^2/m_q$ actually diverges in the chiral limit!
Many quenched simulations actually search for these ``quenched chiral logarithms,''
and a few\cite{QCL}
claim to have seen them.

Quenched QCD will not go away any time soon,
but its days as a source of precision numbers for QCD matrix elements are clearly
numbered. It will continue to be used for tests of methodology, as a proving
ground
for new ways of processing data, and a way to settle some controversies. There is
also quite a bit of continuum phenomenology which can be altered to apply to quenched QCD
and which simulations of quenched QCD can validate -- or not\cite{Faccioli:2003qz}.

\begin{figure}[thb]
\centerline{\psfig{file=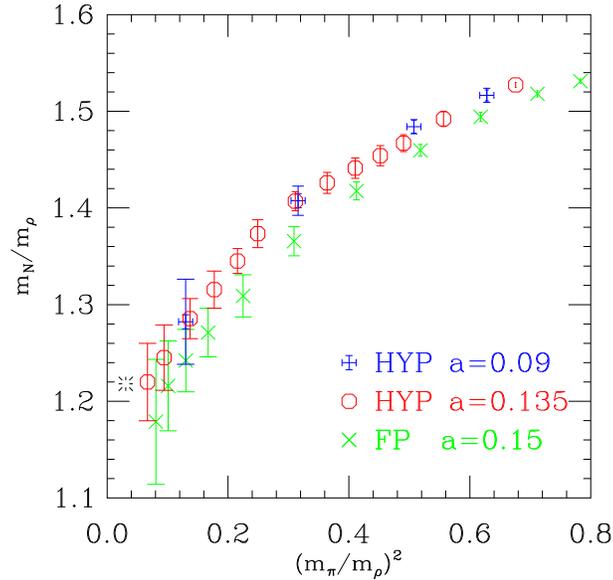,width=8cm}}
\caption{
A comparison of staggered spectroscopy, using an improved action, (pluses and octagons)
at two lattice spacings (from Ref. \protect\refcite{ref:annanpe}) with
that from an improved Wilson-type action (from Ref. \protect\refcite{Gattringer:2003qx}) in crosses.
}
\label{fig:annaape}
\end{figure}

One example of a controversy quenched QCD can address is the following:
At Lattice 2000, S. Aoki\cite{Aoki:2000kp} argued that the continuum limit
of spectroscopy with staggered quarks and Wilson-type quarks might be different. This would
represent a loss of universality, and would represent a serious
problem for (at least) one kind
of lattice discretization of fermion. The data he showed used unimproved Wilson 
fermions (which have order $a$ lattice artifacts), unimproved
staggered quarks, and clover quarks (both of which have $O(a^2)$ scaling violations).
His observed disagreement may have been because of the large lattice
artifacts in (at least some) of the data sets, which make extrapolations
to $a\rightarrow 0$ difficult. An example of more recent data is shown in Fig. \ref{fig:annaape}.
This is an example of an ``APE plot'' which compares two dimensionless ratios
(in this case $(m_{PS}/m_V)^2$ vs $m_N/m_V$). If scaling violations are absent, 
data from
different lattice spacings will lie on a common curve. The data shown in Fig. \ref{fig:annaape}
are not extrapolated to the continuum limit, but  they do give reasonable evidence
that there is not a lot of scale violation in quenched spectroscopy done with improved actions.

\section{Simulations with Dynamical Fermions}

Simulations which include dynamical fermions go back about fifteen years. Because they are so 
costly, people begin with fairly heavy sea quarks
 (pseudoscalar/vector meson mass ratios $> 0.7$), and try to go down. Volumes tend
to be smaller, and lattice spacings tend to be coarser than quenched simulations. 
Statistics are a problem; the simulations have long autocorrelation times.

Over the years
I can recall simulations with 2, 3, and 4 flavors of fermions, using staggered, Wilson, 
and clover quarks.
The RBC collaboration is in the earliest stage of a dynamical fermion
simulation of domain wall fermions\cite{Dawson:2003ph}.
The first simulations with overlap fermions have just appeared: they cost
easily a factor of a hundred times
ones with Wilson quarks\cite{Fodor:2003bh}.
The lattice conferences have had a steady diet
of spectroscopy and matrix element calculations from these projects. But this
year there seemed to be a pause. I think many people\cite{Kaneko:2003re} have decided that
they just cannot push the quark masses down far enough to be interesting,
and have gone back to studying algorithms\cite{Sommer:2003ne}.
(I expect to get a lot of unhappy mail about this sentence.)

\begin{figure}[thb]
\centerline{\psfig{file=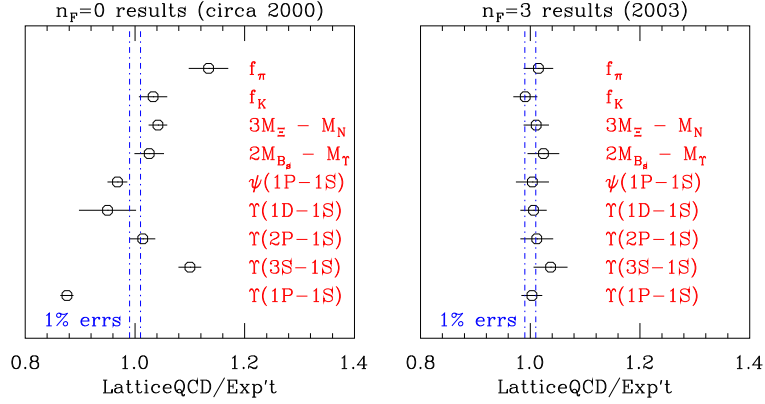,width=10cm}}
\caption{
Comparison of quenched results with results from simulations with 2+1 flavors
of staggered fermions, from Ref. \protect\refcite{Davies:2003ik}.
}
\label{fig:golden}
\end{figure}

There is one notable exception:
The most interesting lattice calculations with dynamical simulations are
the ones being done by the MILC collaboration. They have combined with other groups
to do calculations of spectroscopy with three flavors of dynamical staggered
 quarks, a strange
quark at approximately its correct physical value, and degenerate up and down quarks
whose masses run down to $m_s/5$. Their gauge configurations are used as backgrounds for
simulations with heavy flavors. The analysis uses the taste-breaking chiral Lagrangian
described in Sec. 3.
They have data at two lattice spacings, 0.13 and 0.09 fm, large volumes,
and high statistics. They have studied a variety of processes for which lattice methods
ought to work well: spectroscopy of hadronic states which are not close to
decay thresholds, properties of pseudoscalar mesons, hadrons with
strange valence quarks.
 Their recent preprint\cite{Davies:2003ik}
presents a remarkable agreement of simulation with experiment, shown in
Fig. \ref{fig:golden}.

The MILC part of the staggered project is several years old.
It has produced a lot of interesting spectroscopy. One can clearly
see the effects of sea quarks in the data. Here are some examples\cite{Bernard:2001av}.

\begin{figure}[tbp]
\centerline{\psfig{file=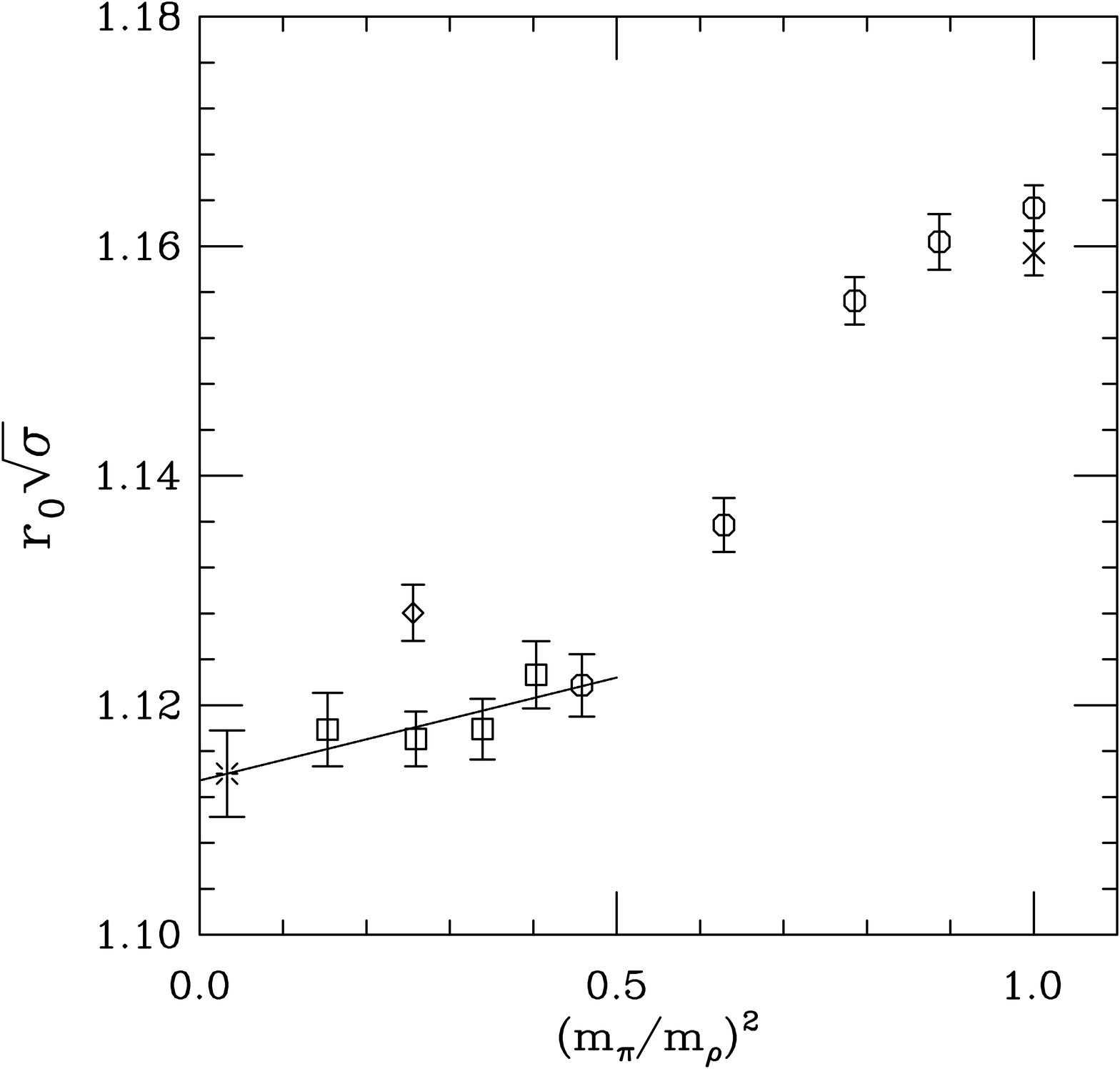,width=8cm}}
\caption{
\label{R0_SIGMA_FIG}
Effects of dynamical quarks on the shape of the potential:
$r_0 \sqrt{\sigma}$ as a function of the quark mass.  The two quenched points
are at the far right, with the octagon coming from the  0.13 fm run and
the cross from the 0.09 fm run.
The remaining octagons are
full QCD runs with three degenerate flavors, and the squares are full
QCD runs with two light flavors and one heavy.
The diamond is the two flavor run, and the burst at the left is a linear
extrapolation of the $2+1$ results to the physical value of $(m_\pi/m_\rho)^2$.
From Ref. \protect\refcite{Bernard:2001av}.
}
\end{figure}

The heavy quark potential in QCD looks something like $V(r) \simeq \sigma r - C/r$.
The location of the bend in the potential is related to a quantity called the Sommer parameter $r_0$
(see Ref. \refcite{ref:precis}) via the force, $-r^2F(r)=1.65$ at $r=r_0$. This corresponds
to $r_0$ of about 0.5 fm.
 Fig.~\ref{R0_SIGMA_FIG}  shows the dimensionless
quantity $r_0\sqrt{\sigma}$ as a function
of the quark mass, represented by $(m_\pi/m_\rho)^2$.  This places the quenched
approximation at $(m_\pi/m_\rho)^2=1$, and the chiral limit at the left side of
the graph.  In these plots the octagons are runs with three degenerate sea quarks,
except for the rightmost point which is the quenched limit. Squares
are runs with $am_s=0.05$ at $a=0.13$ fm. This is roughly the physical value of
the strange quark mass.
For these runs,
$am_{u,d} < 0.05$.  The isolated diamond is a  two flavor
run.  Finally, the cross at $(m_\pi/m_\rho)^2=1$ is an 0.09 fm lattice spacing 
quenched run.
From the two quenched points the authors infer that remaining lattice artifacts are small
compared with the effects of the sea quarks.
One can clearly see the distinction between two and
three flavors, as well as the effect of using two light and one heavy flavor rather
than three degenerate flavors (the ``kink' at $(m_\pi/m_\rho)^2 \approx 0.46$).

Lacock and Michael\cite{UKQCD_J} have observed differences
between the quenched meson spectrum and the real world.  They studied the quantity
\bee J = m_{K^*} \frac{\partial m_V}{\partial m_{PS}^2} \ \ \ ,\ee
where $m_V$ and $m_{PS}$ are the vector and pseudoscalar meson masses.
This quantity has the advantage of being relatively insensitive to the
quark masses, so that accurate tuning of the strange quark mass or
extrapolation of the masses to the chiral limit is not essential.

\begin{figure}[tbp]
\centerline{\psfig{file=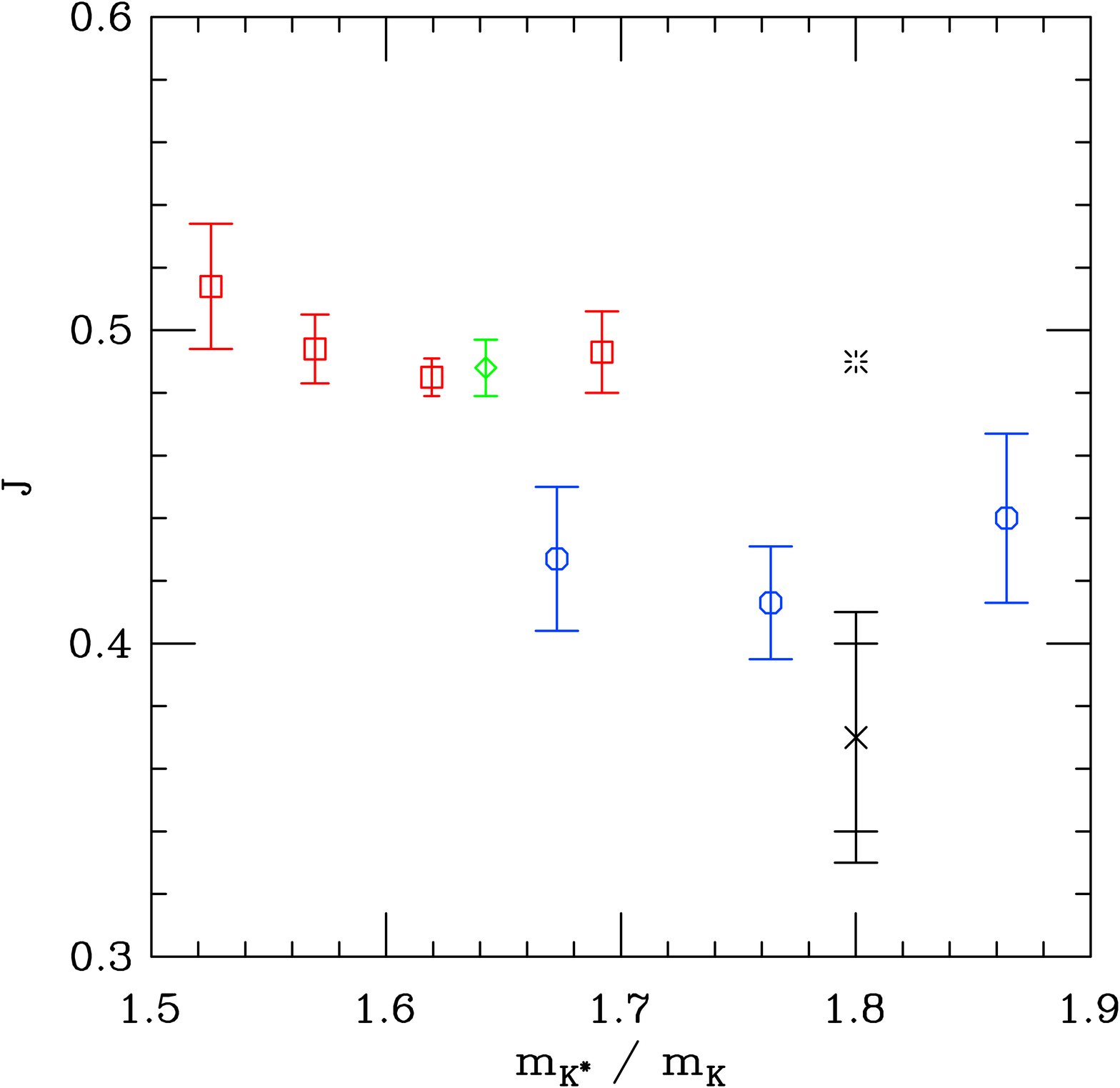,width=8cm}}
\caption{
\label{J_FIG}
The mass ratio ``$J$'' in the quenched and full QCD calculations.
Squares are the three flavor results, and octagons are the quenched results.
The diamond is the two flavor run, using a non-dynamical strange quark
with mass $am_q=0.05$.
The burst is the real world value, and the cross is the quenched
value of Ref. \protect\refcite{UKQCD_J}.
 The smaller error bar on the cross is the statistical error, and
the larger the quoted systematic error. From Ref. \protect\refcite{Bernard:2001av}.
}
\end{figure}

Of course,
to compare to experiment the derivative in this expression must be replaced
by a ratio of mass differences, and MILC choose
\bee \label{J_EQ}  J = m_{K^*} \frac{ \{ m_\phi-m_\rho \}}{2 \{ m_K^2-m_\pi^2 \}} \ \ \ .\ee
Here $m_\rho$ is the mass of the vector meson including two light quarks, etc.
Figure~\ref{J_FIG} shows the results for $J$ in quenched and three flavor QCD.
This is plotted versus $m_{K^*}/m_K$, for which the real world
value is 1.8.  The burst is the real world value of
this definition of $J$ (0.49), and the cross is the value of $J$ found in the
 quenched simulations of Ref. \refcite{UKQCD_J}.  
One can  see a clear effect of the sea quarks on this
quantity.
Figure~\ref{J_FIG} also contains one point with two dynamical flavors.  This
point falls near the three flavor points, indicating that the dynamical
strange quark is less important than the two light quarks.

If there are dynamical fermions in the configurations, then one ought
to see processes like particle decay in the simulations. What the lattice measures is
energies. If a particle $A$ could decay into a $BC$ state, the $A-A$ correlator
 would have a contribution which would show an $\exp(-E_{BC}t)$ behavior, with
$E_{BC}\simeq m_B+m_C$. $E_A$ and $E_{BC}$ might depend on simulation parameters
(like quark masses) and one might look for avoided level crossings as the
parameters were varied\cite{Luscher:1991cf}.
(Spin models with several species of excitation show this kind
of behavior\cite{Gattringer:1992yz}.)
The chief target is the rho meson. Evidence for a $\rho\pi\pi$ coupling has not been seen.
The problem is that the rho is $J=1$, so in the decay $\rho \rightarrow \pi\pi$ the pions are P-waves.
The minimum energy of a two-pion state is then $m_\pi + \sqrt{m_\pi^2 + p_{min}^2}$,
where $p_{min}=2\pi/L$ is the smallest nonzero momentum in a box of width $L$.
This effectively pushes the threshold for the level crossing to smaller quark mass.
(For further discussion of quantum number effects for staggered fermions, see Ref. 
 \refcite{Bernard:an}.)

\begin{figure}[tbp]
\centerline{\psfig{file=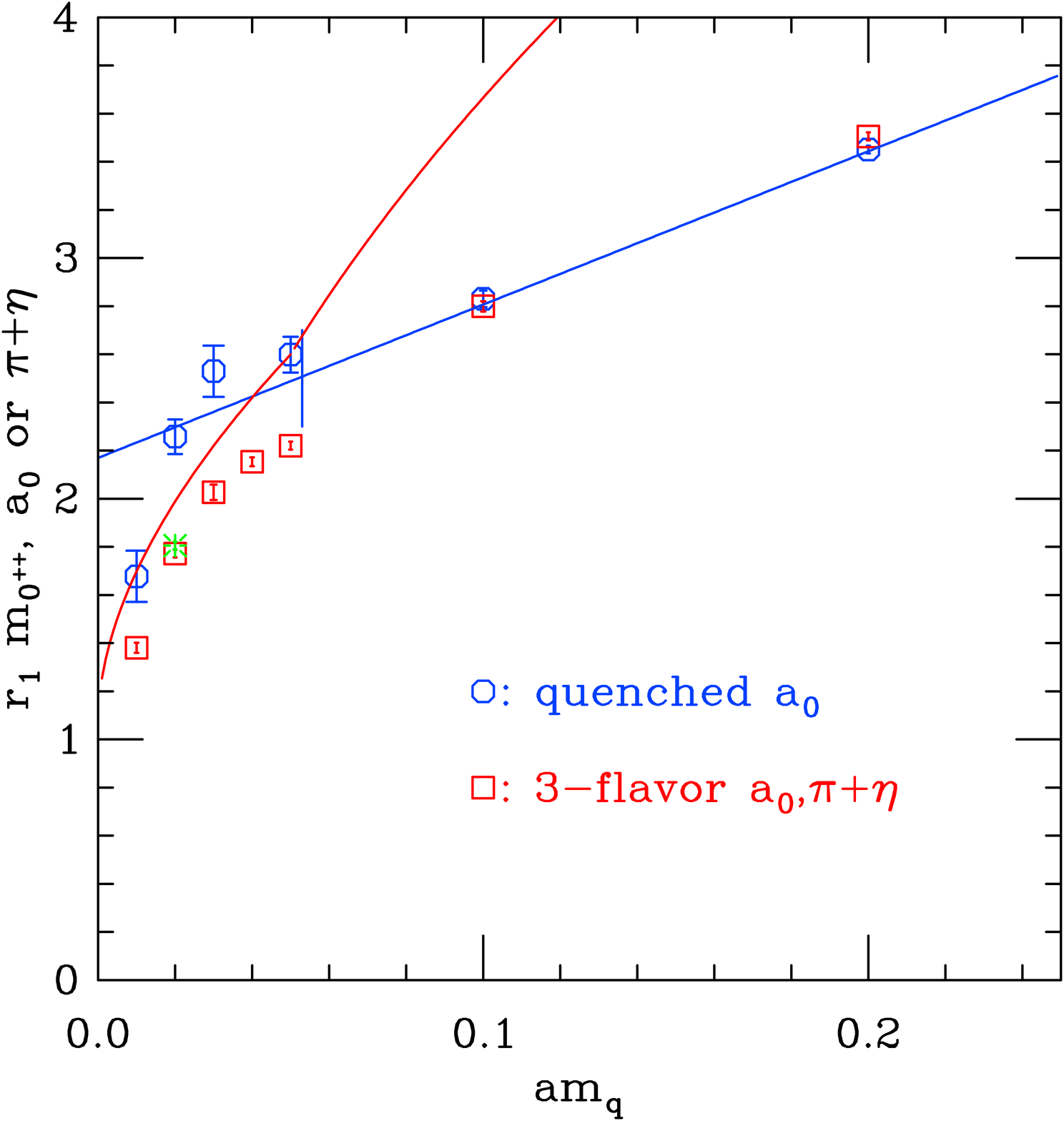,width=8cm}}
\caption{
\label{A0_DECAY_FIG}
$0^{++}$ masses versus quark mass.
The lightest fitted energy in the scalar channel.  Octagons are
quenched results, squares are three flavor results, and the burst
is the two flavor run.
The straight line is a crude extrapolation of the heavy quark points.
The curved line is the $\pi + \eta$ mass estimate, as discussed in the
text.
The short vertical line marks the approximate quark mass where
the $a_0$ mass is twice the quenched pion mass. From Ref. \protect\refcite{Bernard:2001av}.
}
\end{figure}

However, the instability of an excited state has been seen in a different channel,
the isotriplet scalar $a_0$ meson\cite{Bernard:2001av}. It
 is clearly very different in the quenched and
full QCD runs.  For large quark masses there is no visible difference, but as
the quark mass is decreased the full QCD $0^{++}$ mass drops below
all the other masses.  For all but the lowest quark mass, the quenched $0^{++}$
is close to the other P-wave meson masses.  It is plausible to
 ascribe the behavior of the
full QCD mass to the decay of the $a_0$ into $\pi+\eta$.  (Bose symmetry plus
isospin forbids decay into two pions.)  Figure~\ref{A0_DECAY_FIG} illustrates
this interpretation.  In the figure I plot the quenched and full $0^{++}$
masses versus quark mass.  The straight line in the graph is a fit to the
quenched mass for the heavier quarks, and represents the mass of a $q \bar q$
state.  The curved line with the kink at $am_q=0.05$ represents the mass
of $\pi+\eta$.  For $am_q \ge 0.05$ MILC  used three degenerate quark flavors,
so the $\eta$ and $\pi$ are degenerate and this line is simply twice the pion
mass.  For $am_q<0.05$  one does not have direct information on the $\eta$ mass, so
the Gell-Man--Okubo formula written in terms of an ``unmixed $s\bar s$''
mass (just the pseudoscalar mass at $am_q=0.05$) is used:
\bee m_\eta^2 = ( m_\pi^2 + 2 m_{s \bar s}^2 )/3 .\ee
In the quenched case the $a_0$ mesons can couple to two-meson
states through a "hairpin diagram" on one of the meson lines.
Such diagrams can
behave like powers of $t$ times $e^{-2 m_\pi t}$ and therefore
masquerade as a light $a_0$ when $2 m_\pi < m_{a_0}$.  This may
explain the lightest quark mass quenched point\cite{A0STUFF}.

The authors of Ref. \refcite{Davies:2003ik} argue that their results show that
at last, there is a workable algorithm for simulating QCD with light fermions, and that
its successes mean that the time is ripe to apply it to a wide variety of
lattice matrix element calculations.

However, this conclusion is not universally accepted by the lattice community.
Many people are worried about the way staggered fermions are used to represent 2+1
flavors of sea quarks.

Lattice simulations treat valence quarks and sea quarks differently. Basically, valence
quarks (the ones attached to the external sources) do not have any quantum numbers
(other than their mass and spin). One computes
classes of Feynman diagrams on the lattice, then re-weights them with global symmetry indices
 and bundles them
together. (For example, the same propagators are used for the quark and the antiquark in 
a mass-degenerate meson). In staggered fermions, one uses a single flavor of
staggered fermions, with its four tastes, and computes correlation functions in which
the sources project (nearly) onto the same initial and final taste. The quark
could hop temporarily into a different taste state as it propagates across the lattice
(this would happen by emitting and absorbing hard gluons), but this is just
cutoff scale physics which contributes $O(a^2 g^2)$ scale violations.

But sea quarks are different -- the number of flavors matters. In the absence
of flavor symmetry breaking,
mass-degenerate fermions have identical spectra and
one can simulate $N_f$ flavors of degenerate fermions simply by raising $\det  M$ 
for a single flavor to
the $N_f$ power. But each taste of staggered fermions contributes to
the functional integral like a single flavor of continuum quark. Det $ M$
is the weighting for four continuum flavors. To get the weighting for a single flavor, people
doing simulations with staggered fermions re-weight the determinant to $\det^{1/4}M$
per staggered  flavor.

Are simulations with staggered fermions with $\det^{1/4}M$  fundamentally correct?
Until recently, all the theory of the fractional determinant was a single sentence
in a 1981 paper by Marinari, Parisi, and Rebbi\cite{Marinari:1981qf}: ``On the lattice, this action
($S_G - (1/4){\Tr \ln M)}$ will produce a violation of fundamental axioms, but
we expect the violation to disappear in the continuum limit and then recover
the theory with a simple fermion.'' In the past year, the situation has heated up, although
the literature is still sparse\cite{jansen2003}, probably because there are no well-defined
questions.

It is a peculiar exchange-of-limits question. Any N-flavor action action with flavor symmetry
can be reduced to a one-flavor action by taking the $1/N$th root. But in staggered fermions
taste symmetry
is broken by gluon interactions at nonzero lattice spacing. What does taking the fractional
root before restoring flavor symmetry do?

Here is my own condensation of arguments I have participated in\cite{ARGS}:

The most serious problem could be a loss of locality. Is there a single-flavor (undoubled)
local fermion
action whose determinant is equal to $\det^{1/4}M$ (or equal to it up to cutoff effects)?
If such an action exists, then there are no problems in principle with $\det^{1/4}M$ staggered
fermions. This would certainly be something novel! 
Its chiral symmetry properties would certainly be very different from those of
ordinary staggered fermions. Its chiral charge could not be discrete, because that would
violate the Nielsen-Ninomiya theorem. There is no reason that the action could not be ultra-local, 
but it seems hard to combine an ultra-local action with an unquantized chiral charge. 
It's easier for me to imagine that it would  have couplings falling exponentially with distance.
If such an action does not exist, then  universality is lost. It would 
 not possible to integrate out
short distance physics to leave a long distance
effective action whose couplings flow to a fixed point, because remnants of the original
action will be left behind after each blocking step.

And of course, there are many ways to define matrices whose determinants are equal.
Recall the discussion at the end of the last section about taste constraints.
Det $M$ was converted into $\det(M^\dagger M)^{1/2}$ by deleting fields on half the sites
of the lattice. That
gives a ``square root'' action which is ultra-local, as opposed to working
with an action  $(M^\dagger M)^{1/2}$,
which probably isn't local\cite{jansen2003}!

In the momentum-space decomposition of tastes (Eq. \ref{eq:kspace}), each taste occupies a segment of
the Brillouin zone equivalent to $-\pi/2<p_\mu<\pi/2$ with a free field ($i\gamma_\mu \sin(p_\mu)$
dispersion relation. The Fourier transform of its inverse propagator  is an action
with long range (power law falloff) couplings. In the hypercube decomposition, each
 taste's action (the first term of Eq. \ref{stag2})
 is ultra-local, but tastes still mix for any finite value of $a$ through the second term of
Eq. \ref{stag2}.

It is certainly the case that an action with $\det^{1/4}M$ fermions is not unitarity, in the sense
that absorptive parts of processes are not related to cross sections. This is a generic feature
of any partially-quenched theory, where the actions of valence and sea quarks are different.
Different actions mean different Feynman rules. (Even though there may be no action
corresponding to $\det^{1/4}M$ fermions, one can expand $(1/4)\Tr \ln M$ in a power series in $g$
and construct graphs.) Unitarity relates processes in which quarks are internal
lines (with internal vertices, propagators, etc.) to other processes where the same quarks
appear as external lines. Then they would be valence quarks, with valence Feynman rules.
For staggered fermions, the differences occur in processes where taste is changed.
There are $4$ identical vertices (per staggered flavor) which do not change taste in the determinant,
but only one process which changes taste $i$ to another taste $j$ ($i,j=1\dots 4$).
But taste changing interactions involve hard gluons, with momentum
transfer $q_\mu\simeq \pi/a$, so maybe we are just talking about cutoff effects, again.
 It is also unknown whether there are practical consequences
for Euclidean-space lattice calculations of non-unitarity: unitarity describes
 what happens when scattering amplitudes are rotated from
Euclidean space to Minkowski space.

It is generally believed that ``partially quenched'' QCD is well-enough behaved to make
predictions. Indeed there is an extensive literature\cite{Sharpe:2000bc}
on the use of partially quenched chiral perturbation theory to measure
the parameters of full QCD. If the $\det^{1/4}M$ theory is local, then it
is just some funny kind of
partially quenched QCD\cite{Bar:2002nr}  and again nothing would be wrong in principle.

There are several directions this controversy could play out:

First, someone could show analytically that there is some local action which has the
same determinant as $\det^{1/4}M$, or prove that no such action exists. If such an
action exists, then there is nothing wrong in principle with $\det^{1/4}M$ weighting.
It would be an engineering question to ask whether this formalism is good enough
to produce reliable continuum numbers. If no action exists, then QCD with $\det^{1/4}M$
 weighting
is not in the same universality class as continuum QCD. It would still be the case
that its predictions are quite close to experiment (I am not disputing the results
of Ref. \refcite{Davies:2003ik}) for many quantities, but one might be able to
identify processes where it would fail.
It would be like the quenched approximation all over again, only better:
 fundamentally uncontrolled, phenomenologically successful.
I suppose an intermediate possibility is that for sufficiently smooth background configurations,
$\det^{1/4}M$ could correspond to a local action, but for rough fields it might not.
This is easy to visualize when one thinks of the 
fractional power as basically  counting the
geometric mean of the  different taste eigenmodes: $(\prod_i^N \lambda_j)^{1/N}$. At sufficiently
rough coupling, where the chiral symmetry is only a $U(1)$, the eigenmodes are no longer
even approximately degenerate, and the geometric mean would not have much meaning.

Second, one could continue to compute known quantities, see that the simulations
continued to agree with experiment, and go on to do make yet more predictions.
This is  how comparison of theory and experiment generally happens.
The dangerous possibility here is that something might not agree with experiment,
and it might be a matrix element from a Standard Model test. Then, is the
Standard Model wrong, or is it the simulation with a fractional power?

Checks with simpler models would be very useful. The Schwinger model (for different flavor
number) is a good test case, and it is simple enough that it can be overwhelmed by
numerical simulations. Recently, D\"urr and Hoelbling\cite{Durr:2003xs}
have begun a study of the Schwinger model, comparing overlap and staggered fermions.
They have so far looked at only one lattice spacing, and so far
only looked at  condensate- and  topological charge-related quantities.
Results from overlap and staggered fermions are quantitatively different. However,
agreement between staggered and overlap simulations is considerably improved by
smearing the action. This is true both for $N_f=1$ and 2. There does
not seem to be anything funny happening with the $N_f=1$ fractional root.
It may just be that the differences between staggered and overlap fermions are 
cutoff effects, nothing more.
The authors caution that their gauge configurations are much smoother than (my reading)
typical  $a=0.13$ fm four-dimensional $N_f=2+1$ Asqtad  $SU(3)$ configurations, so one should be
cautious with interpretations.
More work needs to be done (and they will do it),
 varying the lattice spacing and checking scaling.

All dynamical fermion simulations are costly -- at least a factor of 100 more than
a quenched calculation done at similar values of lattice spacing and quark mass.
And yet, looking back at Fig. \ref{fig:golden}, the change  in the numbers which come
out of the analyses, going from quenched to 2 + 1 flavor QCD,  is only a few percent.
True, these are quantities which are selected to be insensitive to hadronic decay, or indeed
any strong non quenching effects, but it still seems absurd -- to work so hard for
so little effect. In any classroom physics lecture, this  discussion would be an introduction
to a description of how to do a perturbative calculation, systematically improving
one's approximate result.
But no such story exists for simulations of QCD\cite{DON}.

\section{Hadronic Matrix Elements  from the Lattice }

One of the major goals of lattice calculations is to provide hadronic
matrix elements which either test QCD or can be used to provide low
energy hadronic matrix elements as inputs  to
test the standard model. A wide variety of matrix elements can be computed
on the lattice (compare Fig. \ref{fig:ckm-fig}) and used to extract CKM
mixing angles from experimental data. A new organization, the
Lattice Data Group, {\tt {http://www.cpt.univ-mrs.fr/ldg/}}, is attempting
to build a ``lattice wallet card'' of relevant results.

\begin{figure}[b]
        \bee
        \left(
        \begin{array}{ccc}
        { \bm{V_{ud}}}= 1-{\lambda^2 \over 2}   &  \bm{ V_{us}}=\lambda  &   \bm{ V_{ub} }=A\lambda^3(\rho-i\eta)\\
        \pi\to l\nu & K\to\pi l\nu  & B\to\pi l\nu \\
        \bm{ V_{cd} }= -\lambda  &  \bm{ V_{cs}  }= 1-{\lambda^2 \over 2} &   \bm{ V_{cb}}=A\lambda^2 \\
        D\to l\nu & D_s\to l\nu & B\to D l \nu \\
        D\to \pi l\nu & D\to K l\nu  \\
        \bm{ V_{td}}=A\lambda^3(1-\rho-i\eta)  &\bm{ V_{ts}}= -A\lambda^2  & \bm{ V_{tb}}=1 \\
        \langle B_d | \overline{B}_d\rangle &
        \langle B_s | \overline{B}_s\rangle \\
        \end{array}
        \right) \nonumber
        \ee
\caption{\label{fig:ckm-fig}{Matrix elements which can be computed reasonably reliably
with lattice methods, and their impact on the CKM matrix.
 From Ref. \protect\refcite{Davies:2003ik}.}
}
\end{figure}

The lattice enters at a rather late stage
of the calculation. Consider some ``generic'' electroweak process.
Viewed at the shortest distance scale,
a distance $\Delta x \simeq 1/M_W$, quarks emit and absorb $W$ and $Z$ bosons. However,
physical hadrons containing $u$, $d$, $s$, $c$, or $b$ quarks are
big objects, a fraction of a fermi in size, and they cannot ``see''
the highly virtual W and Z bosons. Their interactions are
described by
a low-energy effective field theory valid at scales $\mu$ of a few GeV, which
would be constructed by integrating out short distance physics using the operator
product expansion, combined with the renormalization group.
The effective Hamiltonian basically reduces to a sum of four-fermion
interactions\cite{ref:buras98}.
 For example, single-W exchange processes would be described by
\bee
H_W^{eff} = {G_F \over {\sqrt{2}}}\sum_{i=0}^{10} c_i(\mu) O_i(\mu)
\label{eq:lowe}
\ee
where the $O_i$'s are four fermion operators and the $c_i$'s are (known)
Wilson coefficients. Similarly, the (second-order weak) process
 $\bar B-B$ mixing is parameterized by the
ratio $x_d  = {{(\Delta M)_{b \bar d}}/ {\Gamma_{b \bar d}}}$
\bee
x_d  = 
\tau_{b \bar d}{{G_F^2}\over{6\pi^2}}\eta_{QCD}F\big({{m_t^2}\over{m_W^2}}\big)
 |V_{tb}^*V_{td}|^2 
 b(\mu) \{ {3\over 8}\langle \bar B| \bar b \gamma_\rho(1-\gamma_5) d
\bar b \gamma_\rho(1-\gamma_5) d | B \rangle  \}
\label{BIGEQ}
\ee
This is a typical formula used to relate an experimental number to a Standard
Model prediction. It has a product of factors
from phase space integrals or perturbative QCD calculations, 
a combination of CKM matrix elements (whose determination is presumably the
primary goal), 
followed by a four quark hadronic
matrix element. In order to extract the CKM
matrix element from the measurement of $x_d$ (and its strange partner
$x_s$), we need to know the value of the
 object in the curly brackets, defined as $3/8 M_{bd}$
and  parameterized as $m_B^2 f_{B_d}^2 B_{b_d}$ where $B_{b_d}$ is the
so-called B-parameter, and $f_B$ is the B-meson decay constant,
$\langle 0 | \bar b \gamma_0 \gamma_5 d | B \rangle =f_Bm_B$.
Lattice calculations are the most model-independent way to compute
the decay constants, B-parameters, and their
ratio $\xi = (\fBs\sqrt{\BBshat})/(\fB\sqrt{\BBhat})$. (Hats denote the
renormalization group invariant quantities.)
The B-parameters and decay constants
are scale and prescription-dependent quantities and the lattice-regulated
quantities have to be converted into some continuum regularization scheme
to complete the calculation.

The lattice calculation of one of these quantities involves many ingredients.
First, one has to extract the lattice-regulated matrix element.
Most of the matrix elements measured on the lattice are
extracted from expectation values of
local operators $J(x)$ composed of quark and gluon fields.
 For example, if one wanted
$\langle 0 | J(x) | h\rangle$ one could look at the two-point function
\bee
C_{JO}(t)= \sum_x \langle 0 | J(x,t) O(0,0) | 0 \rangle .
\label{CURR2}
\ee
Inserting a complete set of correctly normalized momentum eigenstates
\bee
1 = {1 \over L^3} \sum_{A, \vec p}{{|A, \vec p \rangle \langle A, \vec p|}
\over {2E_A(p)}} 
\ee
and using translational invariance and going to large $t$ gives
\bee
C_{JO}(t) = e^{-m_A t} {{\langle 0|J|A\rangle \langle A|O|0\rangle}\over
{2m_A}}. 
\ee
A second calculation of
\bee
C_{OO}(t)= \sum_x \langle 0 | O(x,t) O(0,0) | 0 \rangle
\rightarrow  e^{-m_A t} {{|\langle 0|O|A|\rangle|^2}\over
{2m_A}}
\ee
is needed to extract $\langle 0|J|A\rangle$,  fitting the  two
correlators with three parameters.
 
Similarly, a matrix element $\langle h | J | h' \rangle$ can be gotten
from
\bee
C_{AB}(t,t') = \sum_x \langle 0 | O_A(t) J(x,t') O_B(0) | 0 \rangle.
\label{PROTME}
\ee
by stretching the source and sink operators $O_A$ and $O_B$ far apart
on the lattice, letting the lattice project out the lightest states,
and then measuring and dividing out $\langle 0 | O_A |h\rangle$
and $\langle 0 | O_B|h\rangle $.
 
These lattice matrix elements are not yet the continuum matrix elements.
Typically, one is interested in some matrix element defined with
a particular regularization scheme. It is
 a generic feature of quantum field theory that an operator
defined in one scheme ($\overline{MS}$) will be a superposition of
operators in another scheme (lattice). In principle, the superposition
could be all possible operators. So  an operator of dimension $D$
will mix like
\bee
\langle f | O^{cont}_n(\mu) | i \rangle_{\overline{MS}} =
  a^D\sum_m Z_{nm}\langle f | O^{latt}(a)_m | i \rangle   .
\label{ZFACTOR}
\ee
 The only restrictions on mixing are the ones imposed by symmetries, and generally, lattice
actions have fewer symmetries than in the continuum, so operator mixing is more severe.
 The most serious source of mixing for light quark 
operators is the way lattice fermions treat chiral symmetry. 
Wilson-type fermions break chiral symmetry (even massless ones do so off-shell)
and so nothing prevents mixing into  ``wrong chirality'' operators.
The mixing structure of overlap fermions (and domain wall fermions in the limit of infinite
fifth dimension length)
 is basically identical to the continuum, and 
these formulations really come into their own in cases where chiral symmetry is important.

In Eq. \ref{ZFACTOR} the mixing coefficients to lattice operators with the same dimensionality
as the continuum ones term will contain
the anomalous dimension matrix of the continuum operators
\bee
Z_{mn} = 1 + {g^2 \over{16\pi^2}}(\gamma_{mn} \log \ a\mu + A_{mn}) +\dots
\ee
This must happen to 
 cancel the mu-dependence of the coefficient function, since
$c(\mu)\langle f | O^{cont}|i\rangle_\mu$ is independent of the
 renormalization point. In principle the leading log could be summed,
but in practice people don't know how much of the constant term $A$ should be 
absorbed into a change of scale of $g$. This would involve a two loop
calculation. (Brave groups\cite{Trottier:2003bw} are beginning such calculations.)
So they are just left there as constants.
There are also terms for mixing with higher dimensional operators, which give
contributions proportional to positive powers of $a$. (These are usually
benign, they look just like scaling violations.)
One can also have mixing with lower dimensional operators, with contributions
involving negative powers of $a$.
When they appear, these are dangerous. They must drop out in the
continuum but it is a delicate business, since they look like they are
growing as an inverse power of $a$.

\begin{figure}[thb]
\centerline{\psfig{file=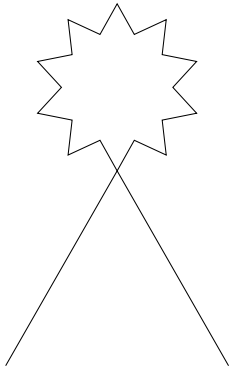,width=3cm}}
\caption{The ``tadpole graph'' for the fermion self energy..}
\label{fig:tadpole}
\end{figure}

People compute the  $Z_{nm}$'s in a number
 of ways. Most straightforward is to use  perturbation theory.
For lattice actions involving thin links,
lattice perturbation theory in terms of the bare coupling $g(a)$
is not very convergent: the $A_{nm}$'s can be very large (order 30, sometimes),
so $\alpha_s A/(4\pi) \simeq 0.2$ or more.
The source of this behavior
 is the ``tadpole graph,'' usually as part of the fermion self-energy.
(See Fig. \ref{fig:tadpole}.)
The lattice fermion-gauge field interaction looks like
$\bar \psi(x) U_\mu(x) \psi(x+\hat \mu)$ and $U \simeq 1 +iga A_\mu - g^2a^2/2
A_\mu^2 + \dots$. The $\bar \psi A_\mu^2 \psi$ vertex, not present in
any sensible continuum regularization, causes problems when the gluon
forms a loop: the quadratic divergence from the loop integral
 combines with the $a^2$ to give
a finite contribution--in fact, it is often the dominant contribution.
In perturbation theory one must also choose the momentum scale in the
running coupling constant. There are reasonable choices for how to do 
that\cite{BLM,PETERPAUL}.
Perturbation theory for fat link actions is generally better behaved.
The fat link vertex acts as a form factor to suppress the tadpole.
Situations where  $|Z|-1$ equals 0.05 or less are not unusual\cite{fatlinkPT}.

Often one can find $Z_{nm}$'s by computing  lattice  Ward
identities\cite{MM}. For example, with overlap fermions, chiral symmetry
``protects'' the combination $m_q \bar \psi \psi$ and constrains the
(local) pseudoscalar and scalar currents, and vector and axial vector currents,
 to have equal renormalization constants. The matrix element of the pseudoscalar
current gives
$\langle 0 | \bar \psi \gamma_5 \psi |PS \rangle= f_{PS}^Pm_{PS}^2/(2m_q)$,
with no lattice-to-continuum renormalization factor for $f_{PS}^P$;
$f_{PS}^P=f_{PS}$.
The zeroth component of the axial current has as its matrix element
$\langle 0 | \bar \psi \gamma_0 \gamma_5 \psi |PS \rangle= f_{PS}^Am_{PS}$,
with $Z_A f_{PS}^A =  f_{PS}$. Thus $Z_A=f_{PS}^P/f_{PS}^A$ which can be computed
by fitting a correlator with a  pseudoscalar
sink and a correlator with an axial vector sink, and taking the ratio.
(To experts, this is slightly wrong: $m_q \bar \psi (1+aD)\psi$ and the
point split currents $\bar \Gamma(1+aD)\psi$ are the quantities related by
overlap current algebra. Experts also know that the ``$aD$'' terms are
removed by a slight redefinition of the lattice propagator...)

 The most commonly used nonperturbative method for computing matching
factors is called the ``Regularization Independent'' or RI scheme\cite{MART}.
One computes quark and gluon Green's functions in a smooth gauge,
regulated by giving all external lines a large
Euclidean squared momentum $p^2$.
For example, for a bilinear, one might compute an un-amputated correlation function
$G_O(p.a)$.
One would also compute the quark propagator in momentum space,
$S(p,a)$. Then the vertex would be found by snipping the
propagators off the correlator, $\Lambda_O(p,a) = S^{-1}(p,a)G_O(p,a) S^{-1}(p,a)$
and the vertex would be computed by projection: $\Gamma_O(p,a) \propto {\rm Tr}\Lambda_O P_O$.
Roughly speaking, the $Z$ factor is the ratio of $\Gamma_O(p,a)$ to its free-field
value, or one would match to a perturbative calculation of the vertex, regulated
identically. To measure a $Z$-factor, one should see
a plateau in the observable versus $p^2$. One must
stay away from
large $pa$ (where discretization effects enter) and small $p$ (where
nonperturbative physics turns on),  The author has not tried doing this himself,
but there are many claims in the literature from people who have, that it
is straightforward to do.

People become very dogmatic about how to compute matching factors. It is a very
interesting exercise to compare the points of view of a firm advocate of lattice
perturbation theory\cite{Trottier:2003bw}
with that of an equally firm advocate of nonperturbative renormalization\cite{Sommer:2002en}.
Read these papers back to back!

\subsection{Charm and bottom matrix elements}
We spent a long section on relativistic fermions, so the reader will not be surprised
to learn that three are also many choices for discretizing heavy fermions.
 It is very difficult\cite{deDivitiis:2003iy} to do direct calculations with relativistic $b$ quarks
because the lattice spacing is much greater than the $b$ quark's
Compton wavelength (or the UV cutoff is below $m_b$). However, it might be
 better
to think of the lattice theory as an effective field theory for 
 low-momentum physics in the presence of two  high
energy scales--the cutoff (inverse lattice spacing) and the heavy quark mass.
The effects of the short distance
are lumped into coefficients of the effective theory
\cite{FERMILAB}.
A simple example of such a theory would be  the standard clover action,
since it has lattice versions of every operator with dimension $\le$ 5.
So the simulation involves a heavy clover quark and (perhaps) some other type
of light quark. The heavy quark dispersion relation is 
 $E(p) \simeq m_1 + p^2/(2m_2)+ \dots$ (if $p<<m$) and it might happen
that $m_1 \ne m_2$ because of discretization artifacts. The heavy  quark magnetic
moment is $\mu=1/m_3$ and again,  $m_3 \ne m_2 \ne m_1$. But $m_1$ is
just an overall constant, so one could shift the measured hadron
mass by $m_2-m_1$ and tune the shifted mass to the desired value (to do
the simulation at the physical $B$ mass, for example).
The lattice magnetic moment could also be tuned (by varying the coefficient of the clover term)
 using (say) the
$B^*-B$ hyperfine splitting. Hadronic parameters not used to determine 
the lattice parameters are then objects of calculation.

Nonrelativistic QCD has also been discretized and used to make very
 precise calculations of the properties of quarkonia \cite{REF6}.
 This formalism can
also be used for the heavy quark  in a B or D meson
(again as long as its momentum is small.)
The ``static'' limit (infinite $b$-quark mass)
 is often used as an additional point on the
curve.  
 Then one can  try to interpolate all the way from light quarks to heavies
and get all the decay constants at once. 

Another new development is the use of staggered quarks for the light
quark in a heavy-light system\cite{Wingate:2002fh}. This will allow one to
get to smaller quark masses than Wilson-type valence quarks permit, which
will be important for chiral extrapolations. Overlap light quarks
are even prettier\cite{Becirevic:2003hd}
but much more costly. Calculations using overlap fermions will probably appear next year.

Early form of fat links failed rather spectacularly\cite{Bernard:2000ki,BERNARD}
 when used with heavy flavor
simulations
because they smeared away the short distance part of gluonic interactions,
but HYP links have been used to enhance the signal of static-light
 correlators\cite{DellaMorte:2003mw}.

An important systematic effect is the use of unphysical (too heavy) values of
the light quark masses both in quenched and unquenched
simulations.  
People had just been doing linear extrapolations 
to get down to the physical light quark mass,
 but this is not correct\cite{kr_02,panel_lat02,slov_02}!
One has to use chiral perturbation theory, with its non-analytic (logarithmic)
behavior at small $m_q$.
An example of such behavior is
\beea
   & & {{\fBs\over\fB}}-1 = (m_K^2-m_\pi^2){f_2(\Lambda)}  \nonumber \\
   & & -{{1+3{g^2}}\over{(4\pi f_\pi)^2}} \left[
         \frac{1}{2}I(m_K)+\frac{1}{4}I(m_\eta)
        -\frac{3}{4}I(m_\pi) \right]
\eea
where $I_{\rm P}(m_{PS})=m_{PS}^2\ln(m_{PS}^2/\Lambda^2)$ and $f_2$ is a
low-energy constant. Kronfeld and  Ryan\cite{kr_02} pointed out that
 the inclusion of chiral logarithms in the chiral
extrapolation of lattice data for heavy-light decay constants can
drastically change $\fBs/\fB$ and  $\xi$. By assuming
$g^2=g_{D^*D\pi}^2=0.35$ \cite{Anastassov:2001cw}
 and $f_2=0.5(3)$ GeV${}^{-2}$,
they concluded that $\xi=1.32\pm 0.10$, which is more than
10 per cent larger than Ryan's 2001  global lattice estimate\cite{Ryan:2001ej}.

Most lattice results are done in quenched approximation. In many cases,
all the other systematics (extrapolation to the continuum, matching
factors, etc.) can be beaten down to below
a few per cent. But the quenching systematic won't go away until
dynamical fermion simulations are done.

There are many lattice calculations of $f_B$, $f_D$, $B_B$, and form factors
for semileptonic decay.
My summary leans  rather heavily on reviews by Wittig\cite{Wittig:2003cf},
Becirevic\cite{Becirevic:2003hf},
and Kronfeld\cite{Kronfeld:2003sd}.

The decay constant is computed by combining a heavy quark and a light antiquark
propagator into Eq. \ref{CURR2}.
Decay constants  probe very simple properties of the wave function: in the
nonrelativistic quark model
$
f_M = {{\psi(0)}/{\sqrt{m_M}}} 
$,
where $\psi(0)$ is the $\bar q q$ wave function at the origin.
For a heavy quark ($Q$) light quark ($q$) system $\psi(0)$ should become
independent of the heavy quark's mass as the $Q$ mass goes to infinity, and
in that limit one can show in QCD that $\sqrt{m_M}f_M$ approaches a constant.

\begin{figure}[thb]
\centerline{\psfig{file=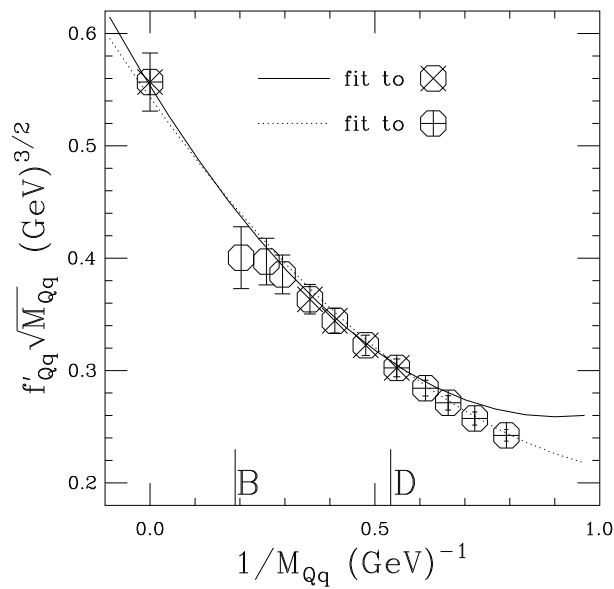,width=8cm}}
\caption{
Pseudoscalar meson decay constant vs $1/M$, from Ref. \protect\refcite{BERNARD}.}
\label{fig:frootm}
\end{figure}

\begin{figure}[thb]
\centerline{\psfig{file=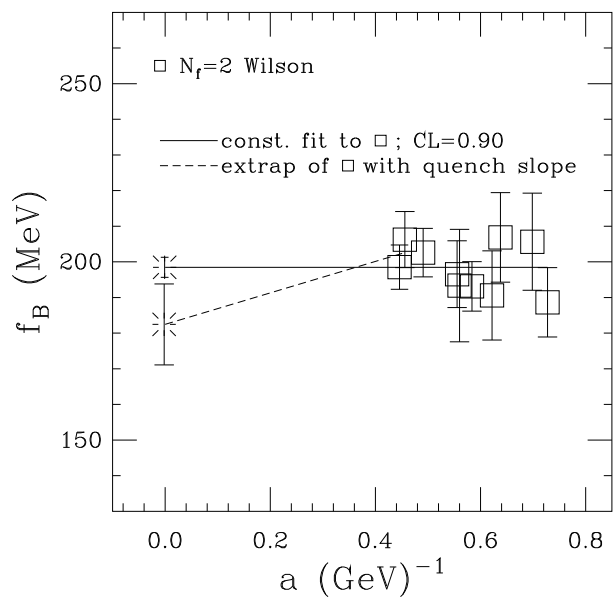,width=8cm}}
\caption{ $f_B $ vs. $a$ from Ref. \protect\refcite{BERNARD}, 
showing extrapolations to
the continuum limit of  full  QCD data.
}
\label{fig:fb}
\end{figure}

The lattice calculation of  decay constants which I know best is by the MILC collaboration,
again,
 Refs.
 \refcite{BERNARD} ($N_f=2$) and \refcite{Bernard:2002ep} ($N_f=2+1$, preliminary results).
These authors 
have done careful  simulations, in quenched, 2-flavor, and 2+1 flavor QCD,
 at many values of the lattice
spacing, which allows one to extrapolate to the continuum limit by brute
force.  Examples of  results are shown in Figs.
\ref{fig:frootm} and \ref{fig:fb}.
The first figure shows $f\sqrt{M}$ vs $1/M$, with data extending from
the static limit to below the $D$. One would fit this data to
a power series in $1/M$ to get to the $B$ or $D$.
Notice that the simple heavy quark assumption $f_M \propto 1/\sqrt{M}$
fails by 50 per cent at the $B$.

A plot of the data which goes into Fig. \ref{fig:frootm} is shown in Fig. \ref{fig:fb}.
There does not seem to be a lot of lattice spacing dependence. Quenched data extends
to smaller lattice spacing, and so one might want to use its slope to extrapolate
to the continuum. The two $a=0$ points show different extrapolation choices,
clearly a source of uncertainty.

\begin{figure}[thb]
\centerline{\psfig{file=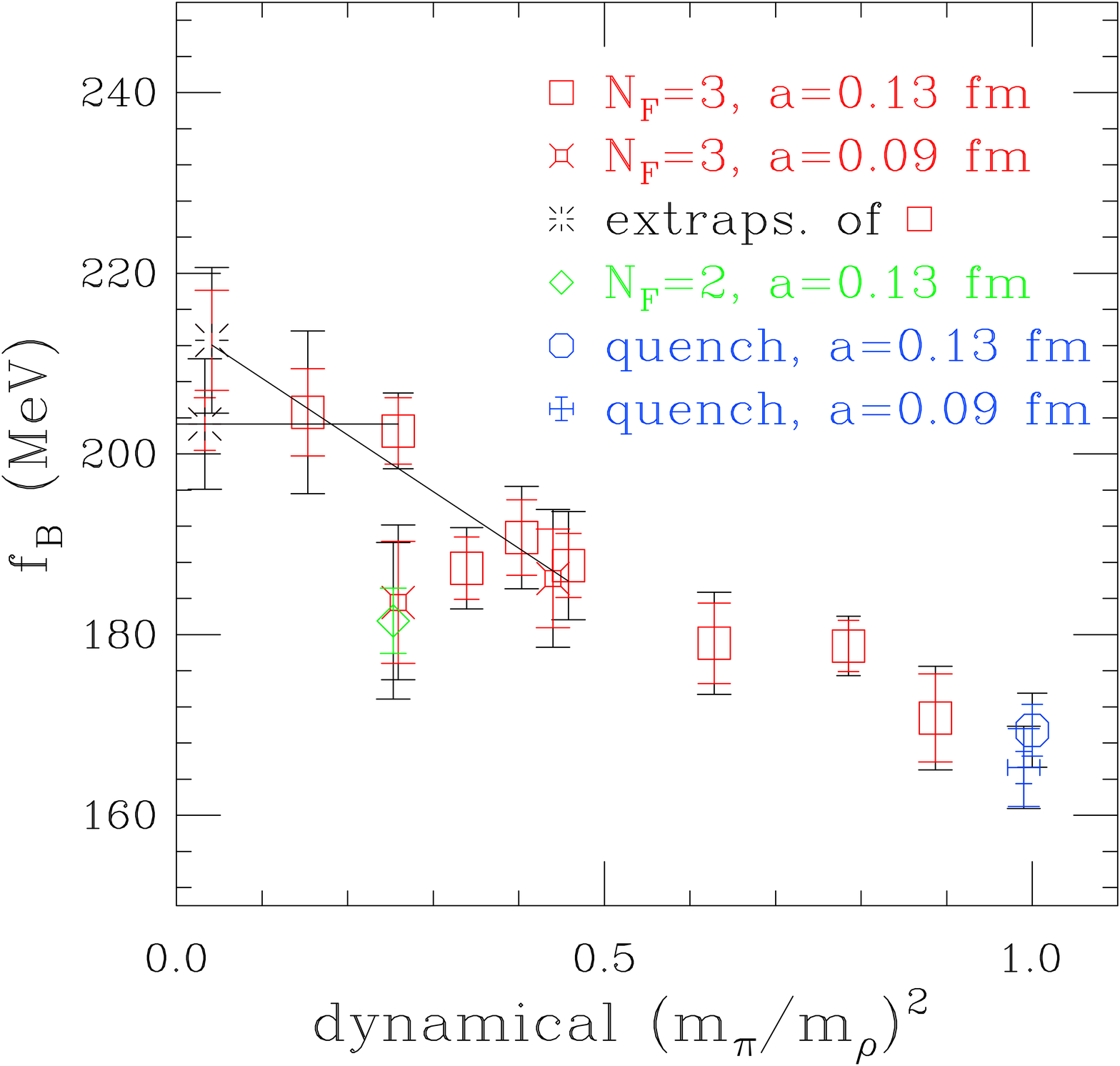,width=8cm}}
\caption{ $f_B $ as a function of dynamical quark mass (plotted as $(m_\pi/m_\rho)^2$) from
 Ref. \protect\refcite{Bernard:2002ep}. There appears to be a considerable shift as the sea quarks
drop in mass. 
}
\label{fig:fbqtod}
\end{figure}

\begin{figure}[thb]
\centerline{\psfig{file=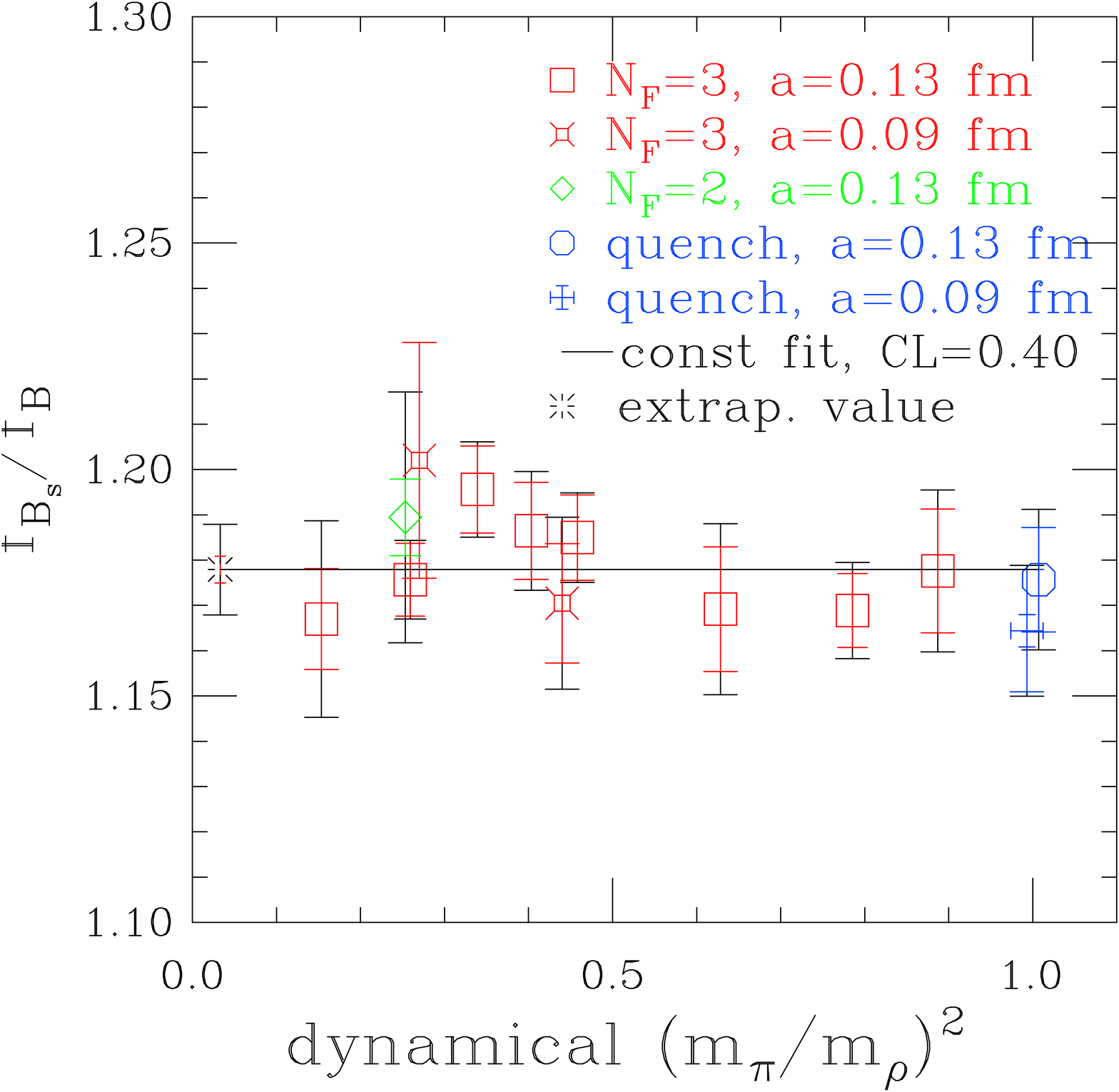,width=8cm}}
\caption{ $f_{B_s}/f_B $ as a function of dynamical quark mass (plotted as $(m_\pi/m_\rho)^2$)
 from Ref. \protect\refcite{Bernard:2002ep}. This ratio is almost independent of
the sea quark mass.
}
\label{fig:fbsfbqd}
\end{figure}

Figure \ref{fig:fbqtod} 
from Ref. \refcite{Bernard:2002ep}
shows how $f_B$ changes as the sea quarks switch on -- a  jump from
170 to 210 MeV. The ratio $f_{B_s}/f_B$, shown in Fig. \ref{fig:fbsfbqd}, seems to be much more stable.
Sometimes, apparently, quenching works well, sometimes not.


Semileptonic decays involve evaluating lattice correlators  of three-point functions
(two mesons and the current). One just fits the data (measured hopefully
at many masses and as many momenta as possible--remember,
momentum is quantized to be a multiple of $2\pi/L$ in a box of size $L$) to
 the expected set of form factors. For example, in
$B\rightarrow \pi \ell \nu$, 
\bee
\langle \pi(p) | V_\mu |B(p')\rangle =
 f^+(q^2)[p'+p - {{m_B^2-m_\pi^2} \over q^2}q]_\mu +
 f^0(q^2){{m_B^2-m_\pi^2} \over q^2}q_\mu .
\ee
 The best signals come when the momenta of the initial
and final hadron are small.
 Then the large $B$ mass forces
$q^2$ ($q={}$lepton 4-momentum${}=p_B-p_\pi$) to be  large.
If the form factor is needed  at 
$q^2\sim0$, a large extrapolation is needed,
and there will be additional errors and model dependence in the answer.
(Lattice people have no advantage over anyone else at guessing 
functional forms.)
  However, finding $V_{ub}$
from experimental data only requires knowing the form
factor at one value of $q^2$. This should work so long as the experiment has
enough data to measure the differential rate 
 around that region of $q^2$.

As an example of a recent approach, the
FNAL group has measured $B\to D\ell\nu$ and  $B\to D^*\ell\nu$ form factors at zero 
recoil\cite{FNALPRD}. They have a clever technique for removing much
of the lattice-to-continuum Z-factors by computing ratios of
matrix elements. For example, in $B\to D\ell\nu$, the differential
cross section at zero recoil  involves a form factor $h_+(v\cdot v'=1)$. 
This quantity is one in the heavy quark limit\cite{Isgur:vq} with corrections\cite {Falk:1992wt}
which
start at $1/m_c^2$. The lattice calculation uses lattice operators
and needs to be matched to a continuum result. The FNAL group extracted it
from the dimensionless ratio
\bee
{{
\langle D | \bar c \gamma_0 b | \bar B \rangle
\langle \bar B | \bar b \gamma_0 c |  D \rangle
}\over{
\langle D | \bar c \gamma_0 c | D \rangle
\langle \bar B | \bar b \gamma_0 b | \bar B \rangle
}} = |h_+(v\cdot v'=1)|^2
\ee
The denominators are just diagonal matrix elements of the charge density,
(i.e. they are basically just charges) and the ratio of renormalizations
between numerator and denominator is under better control than the
individual factors. The $B\to D^*\ell\nu$ calculation is similar. Their calculation
of its form factor is $h(1)=0.913$ with a combined non-quenching error of about two per cent.
Using this number in conjunction with experimental measurements of the branching
ratio would give about a five per cent uncertainty in $|V_{cb}|$, not including
the quenching uncertainty, of course.

\begin{figure}[thb]
\centerline{\psfig{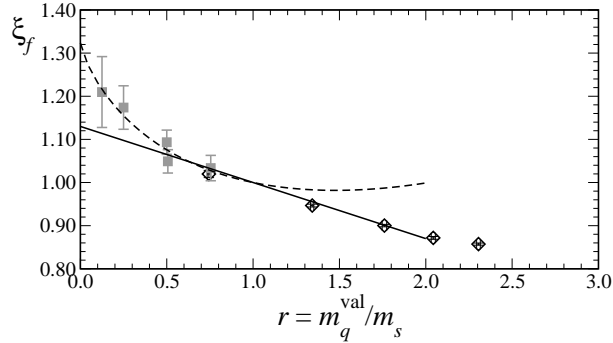}}
\caption{ $\xi$ vs quark mass. The black points are from
Ref. \protect\refcite{Aoki:2003xb} while the grey ones are from
 Ref. \protect\refcite{Wingate:2003ni}. The straight line is a linear extrapolation
and the dotted line is a chiral logarithm fit,
both from Ref. \protect\refcite{Aoki:2003xb}.
This picture is from the review of
Kronfeld\protect\cite{Kronfeld:2003sd}.
}
\label{fig:kronxi}
\end{figure}

Finally, $\xi$. I know of two published calculation of $\xi$ in $N_f=2$ QCD,
which includes chiral logarithms in their analysis. The first is by the 
JLQCD collaboration\cite{Aoki:2003xb}.
Their quark masses ($0.7m_s<m_q<2.9m_s$) are too large to actually see
any chiral logarithm, but they can bound the size of the effect through their inability
to see it.  Recently, Wingate et al\cite{Wingate:2003ni}
presented results (from simulations using Asqtad valence quarks on the MILC
Asqtad background configurations) which are consistent with the behavior expected from
chiral logarithms. A value of $\xi$ well away from unity is suggested, about 1.25.
Kronfeld's summary is shown in Fig. \ref{fig:kronxi}. The light quark
data of Ref. \refcite{Wingate:2003ni}  tracks the fit of Ref. \refcite{Aoki:2003xb}.
However, it happens
that the decay constants of the two groups are
quite different: about 260 MeV for Ref. \refcite{Wingate:2003ni} as opposed to
215 MeV for Ref. \refcite{Aoki:2003xb}. Errors in both cases are claimed
to be about ten per cent from
all sources.  Clearly something is not working... A glance at Fig. \ref{fig:fbqtod} shows
a third  prediction for $f_B$ of about  210 MeV (from $f_B$, see Fig. \ref{fig:fbqtod}), 
 $\times 1.2$ for $\xi$ from Fig. \ref{fig:fbsfbqd} $\simeq 250$ MeV, with maybe
a ten per cent error.

\subsection{Kaon  Matrix Elements}

\subsubsection{$B_K$}
The kaon B-parameter $B_K$,  defined as the matrix element of a particular
four fermion operator $O_+$,
\bee
{8\over 3} (m_K f_K)^2 B_K =\langle \bar K| \bar s \gamma_\mu(1-\gamma_5) d
\bar s \gamma_\mu(1-\gamma_5) d | K \rangle ,
\label{eq:bk}
\ee
 is an important ingredient in the testing
of the unitarity of the Cabibbo-Kobayashi-Maskawa matrix \cite{Hocker:2001xe}.
It has been a target of lattice calculations since the earliest
 days of numerical simulations of QCD. $B_K$ puts a severe test on the chiral
behavior of lattice simulations and so it is a useful test of methodology (in addition
to being a dimensionless number interesting to experiment).
 There has been a continuous  cycle of lattice calculations using
 fermions with ever better chiral properties. Unfortunately, to date
all calculations (except one preliminary one described below)
 have been done in quenched approximation, and
so there is no lattice prediction relevant to experiment.

The first lattice calculations of $B_K$ were done with Wilson-type
actions. These actions break chiral symmetry. This is a problem, because
while the matrix element of $O_+$ scales as $m_{PS}^2$, the opposite
chirality operators which appear under mixing have matrix elements which are
independent of $m_{PS}$. The signal disappears under the background in the chiral limit.
Techniques
for handling operator mixing have improved over the years,
 (for  recent results,
see Refs. \refcite{Gupta:1996yt}-\refcite{Aoki:1999gw})
 but I think this approach is still very difficult.

Staggered fermions (Refs. \refcite{Kilcup:1997ye}--\refcite{Aoki:1997nr})
have enough chiral symmetry at nonzero lattice spacing, that
operator mixing is not a problem. One can obtain extremely precise
 values for lattice-regulated
 $B_K$ at any fixed lattice spacing. However, to date, all calculations
 of $B_K$ done
 with staggered fermions use ``unimproved'' (thin link,
 nearest-neighbor-only interactions),
and scaling violations are seen to be large. For example,
 the JLQCD collaboration\cite{Aoki:1997nr}
saw a thirty per cent variation in $B_K$ over their range
 of lattice spacings. Fat link calculations are in progress\cite{Bhattacharya:2003up}.

For domain wall fermions,   chiral
 symmetry remains approximate, though
much improved in practice compared to Wilson-type fermions. Two 
groups\cite{AliKhan:2001wr,Blum:2001xb} have presented results
 for $B_K$ with domain wall fermions.
Ref. \refcite{AliKhan:2001wr} has data at two lattice
 spacings and sees only small scaling violations.
There is a few standard deviation disagreement
 between the published results of the two groups.

Finally, overlap actions have exact chiral symmetry
 at finite lattice spacing. All operator
mixing is exactly as in the continuum\cite{Capitani:2000da}.  Two
groups\cite{Garron:2003cb,DeGrand:2002xe} have recently
 presented  results for $B_K$.

\begin{figure}[thb]
\centerline{\psfig{file=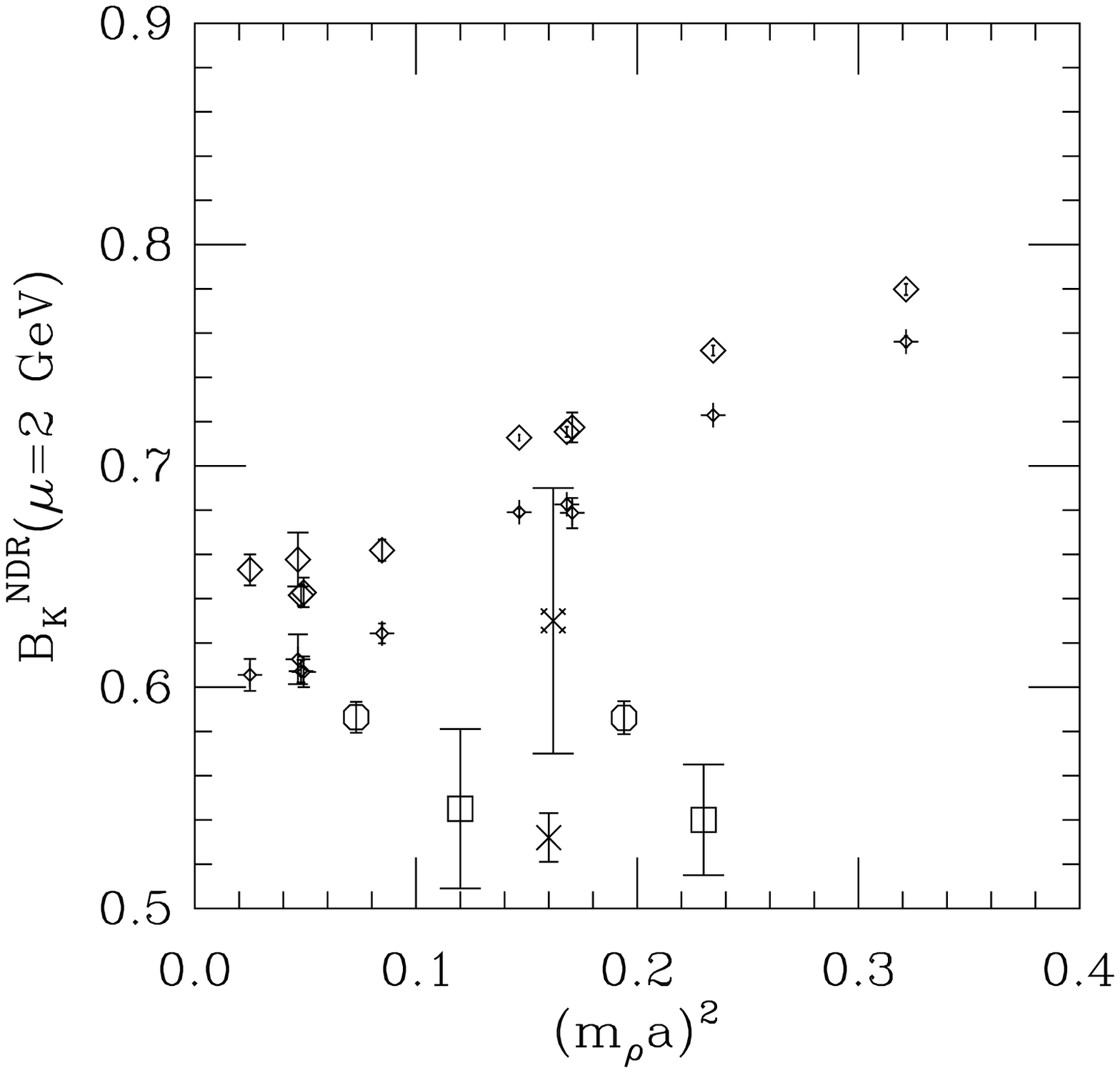,width=8cm}}
\caption{
$B_K$ comparisons vs lattice spacing, from a selection of simulations
with reasonably small error bars. Results are labeled
diamonds and fancy diamond, Ref. \protect\refcite{Aoki:1997nr},
the fancy cross, Ref. \protect\refcite{Garron:2003cb},
octagons, Ref.\protect\refcite{AliKhan:2001wr},
the cross, Ref. \protect\refcite{Blum:2001xb},
and squares, Ref. \protect\refcite{DeGrand:2002xe}.
}
\label{fig:worldbka}
\end{figure}

\begin{figure}[thb]
\centerline{\psfig{file=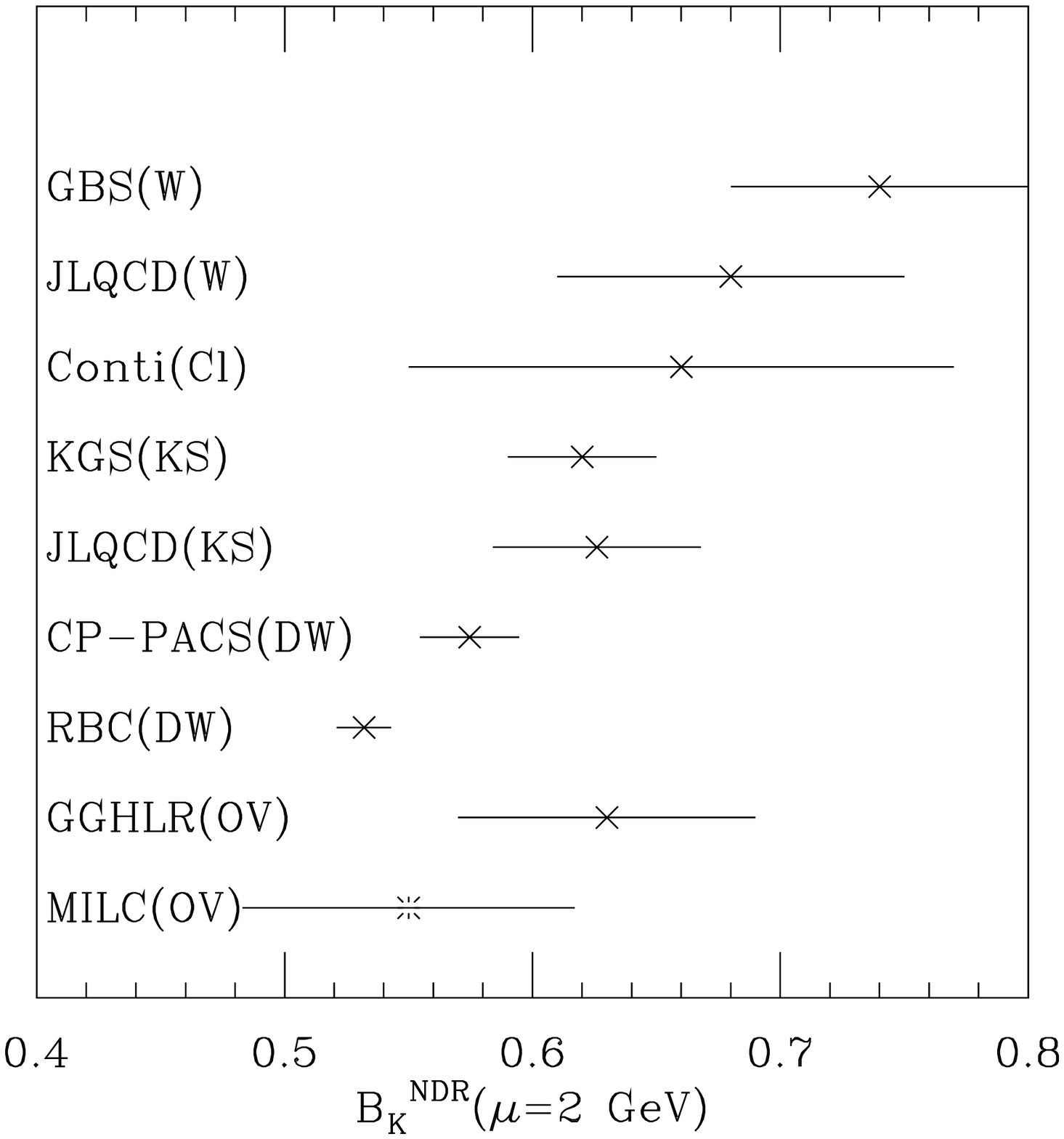,width=8cm}}
\caption{
$B_K$ comparisons presented ``as if'' they were taken to the continuum limit.
The label in parentheses characterizes the kind of lattice fermions used:
W for Wilson, Cl for Clover, KS for staggered, DW for domain wall, and
OV for overlap fermions.
References are
GBS, Ref. {\protect\refcite{Gupta:1996yt}},
JLQCD(W), Ref. {\protect\refcite{Aoki:1999gw}},
Conti, Ref. {\protect\refcite{Conti:1997qk}},
KGS, Ref. {\protect\refcite{Kilcup:1997ye}},
JLQCD(KS), Ref. {\protect\refcite{Aoki:1997nr}},
CP-PACS, Ref. {\protect\refcite{AliKhan:2001wr}},
RBC, Ref. {\protect\refcite{Blum:2001xb}},
GGHLR, Ref. {\protect\refcite{Garron:2003cb}},
and MILC \protect\refcite{DeGrand:2002xe}.
The points of Refs.
{\protect\refcite{Aoki:1999gw}},
{\protect\refcite{Kilcup:1997ye}},
{\protect\refcite{Aoki:1997nr}},
{\protect\refcite{AliKhan:2001wr}},
and \protect\refcite{DeGrand:2002xe}
 are the results of a  a continuum extrapolation; all the rest are
 simulations at one lattice spacing.
}
\label{fig:worldbk}
\end{figure}

Figure \ref{fig:worldbka} shows results for $B_K$ at various lattice spacings,
for a selection of
simulations which have reasonable statistics and small error bars.
Figure \ref{fig:worldbk} presents results which are either extrapolated
to the continuum limit, or presented by their authors as having
 small lattice spacing artifacts.

\begin{figure}[thb]
\centerline{\psfig{file=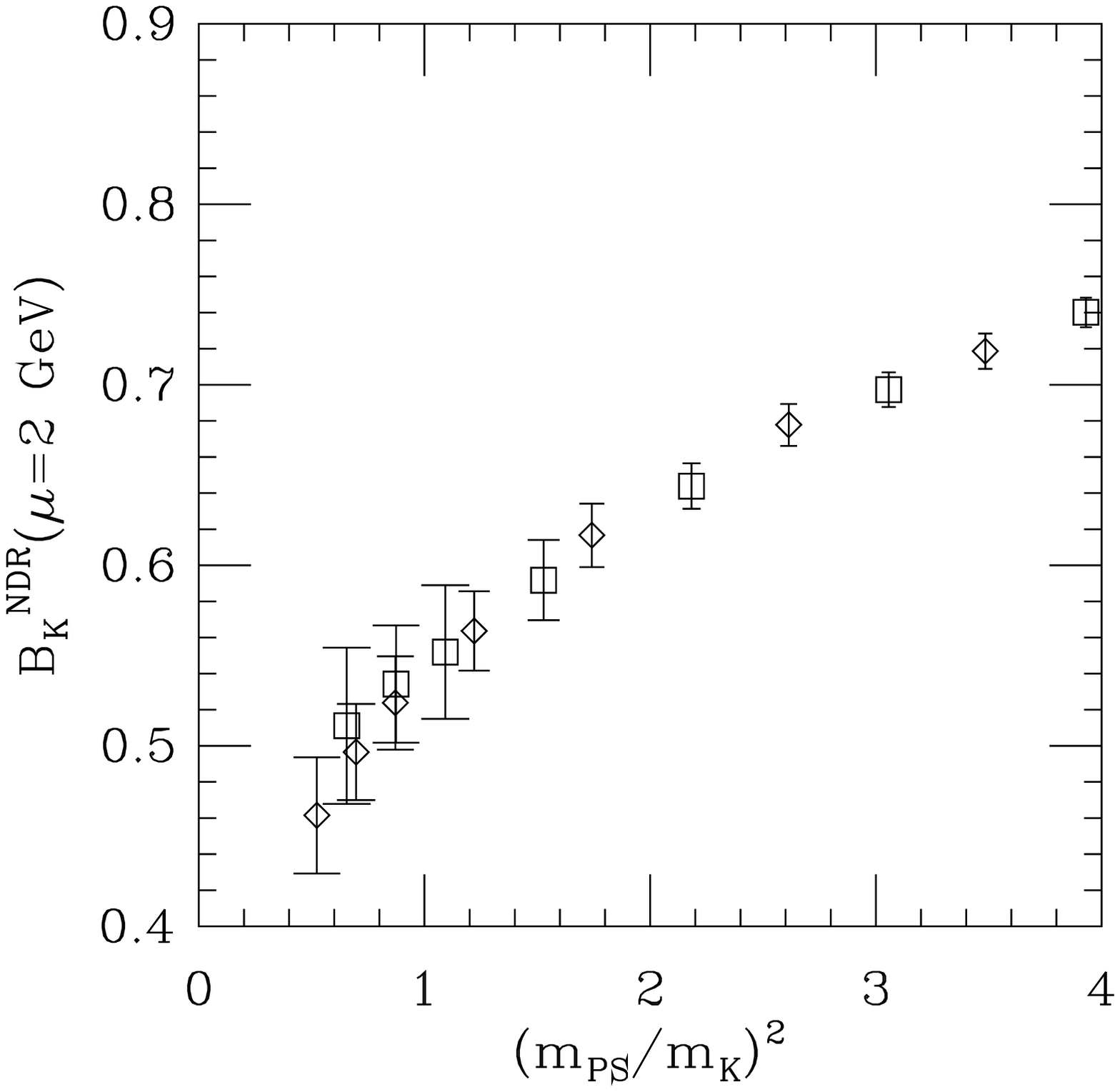,width=8cm}}
\caption{ A typical (not very good) lattice calculation\protect\cite{DeGrand:2002xe}
 of $B_K^{NDR}(\mu=2$ GeV)
 as a function of quark mass from the two data sets--diamonds for $a=0.13$ fm,
squares for $a=0.09$ fm. }
\label{fig:combinedbk}
\end{figure}

$B_K$ shows quite a bit of nonlinear  quark mass dependence (compare Fig. \ref{fig:combinedbk}
for a typical recent result). This suggests strongly that chiral logarithms are present in the
data,
and it is reasonably easy to fit  the data including their effects, out to quark masses which are about 
equal to the strange quark's mass. It happens that the coefficient of the chiral logarithm is the same
in full and quenched QCD.

The only group reporting results for a calculation of $B_K$ in unquenched
 QCD that I am aware of is
the RBC collaboration, using domain wall fermions\cite{Izubuchi:2003rp}.
At this year's lattice conference, they presented a preliminary
number of $B_K^{NDR}(\mu=2 \ {\rm GeV}) =0.50(2)$ for $N_f=2$.
It is low
compared to all the quenched $B_K$ results quoted above.
This is interesting: a too-low value of $B_K$ moves the region
in the $(\rho,\eta)$ plane of the CKM matrix allowed by $B_K$ (actually, $\epsilon_K$) away from the
heavy flavor determinations. The authors do not claim anything so grand; their
calculation is still in its early stages. Phenomenologists should keep an eye on their result,
however.

\subsubsection{$K\rightarrow\pi\pi$ and $\epsilon'/\epsilon$}
There are two types of CP violating decays of kaons. The first (and largest) origin
is the admixture of a small amount of a CP eigenstate into the wave function
of a particle  which is dominantly of the other eigenstate. This is called
``indirect CP violation,'' as opposed to  `direct CP violation,'' in which the decay of the
large component of the wave function  violates CP. The ratio of these two
processes is parameterized by the ratio
\bee
{{\epsilon'}\over \epsilon} = (16.6 \pm 1.6) \times 10^{-4}
\label{eq:exptee}
\ee
where I am quoting its 2003 experimental value, from the review of Buras\cite{Buras:2003wd},
based on experimental data of
 Refs. \refcite{Lai:2001ki,Alavi-Harati:1999xp,Burkhardt:1988yh,Gibbons:zq}.

About two years ago, the RBC\cite{Blum:2001xb} and CP-PACS\cite{Noaki:2001un} collaborations
computed $\epsilon'/\epsilon$ in
quenched approximation, using domain wall fermions. Their numbers are in spectacular
disagreement with Eq. \ref{eq:exptee}: Ref. \refcite{Blum:2001xb} quotes
$-7.7(2) \times 10^{-4}$
and Ref. \refcite{Noaki:2001un} says that their value is ``negative, and has a magnitude
of order $10^{-4}$. What is going on?

Lattice calculations of $\epsilon'/\epsilon$ are complicated. They begin with the low
energy effective Lagrangian of Eq. \ref{eq:lowe} to compute (first) $K\rightarrow\pi\pi$ amplitudes
\bee
K^0 \rightarrow (\pi \pi)_I\,=\,A_I\exp(i\delta_I)
\ee
and the parameter related to the $\Delta I = 1/2$ rule,
\bee
\frac{1}{\omega} = \frac{{\rm Re}(A_0)}{{\rm Re}(A_2)}\approx 22.
\ee
These feed into the expression for $\epsilon'/\epsilon$,
which happens to be dominated by the difference
of two of the operators, the QCD penguin
\bee
Q_6=({\bar s}d)_L \left(
({\bar u} u)_R +({\bar d} d)_R+({\bar s} s)_R\right),
\label{eq:o6}
\ee
and the electroweak penguin
\bee
Q_8=\frac{1}{2}({\bar s}d)_L \left(
(2{\bar u} u)_R -({\bar d} d)_R-({\bar s} s)_R\right),
\label{eq:o8}
\ee
\bee
\frac{\epsilon'}{\epsilon}\approx \frac{\omega G_F}{2|\epsilon|{\rm Re}A_0}
{\rm Im}(V_{ts}V_{td}^*)\times
\label{8.3}
\end{equation}
\begin{equation}
\left(y_6(\mu)\langle Q_6\rangle^{\overline {MS}}_{I=0}(\mu)-\frac{1}{\omega}
y_8(\mu)\langle Q_8\rangle^{\overline {MS}}_{I=2}(\mu)\right) \nonumber
\ee
and
$\langle Q_i\rangle_I\exp(i\delta_I)=\langle \pi \pi_I|Q_i|K\rangle $.

People don't calculate $K\rightarrow \pi\pi$ directly on the
 lattice; it is difficult\cite{MAINIETAL} to extract the phase shifts
from the $\pi\pi$ final state interactions from lattice data (never mind
trying to separate the two pions to asymptotically great distances).
Instead, they use chiral perturbation theory~\cite{CURRAL} to relate
$K\rightarrow \pi\pi$ amplitudes to $K\rightarrow \pi$ and $K\rightarrow$ vacuum.
 In the case of the
$\Delta I =3/2$ amplitude there is a factor of two change in the lattice
result depending on whether tree level or one loop chiral perturbation theory
is used. This is shown in an old figure from Ref. \refcite{KURA} in Fig. \ref{fig:di32}.

\begin{figure}[thb]
\centerline{\psfig{file=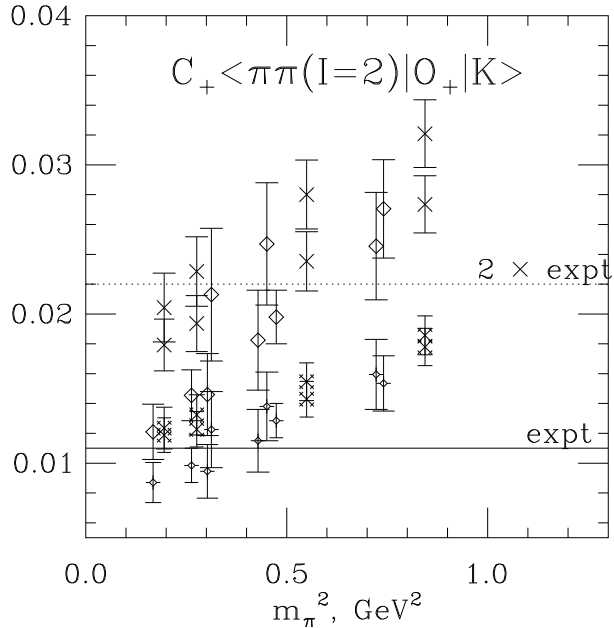,width=8cm}}
\caption{$\Delta I = 3/2$ $K\rightarrow \pi\pi$ amplitude with (fancy symbols)
and without (plain symbols) one-loop corrections of quenched chiral
 perturbation theory. Data are
 crosses and fancy crosses, Ref. \protect\refcite{BERNA};
 diamonds and fancy diamonds, Ref. \protect\refcite{JLDI}.
}
\label{fig:di32}
\end{figure}

At the end of the day, there is also a conversion from the lattice
regulated matrix element to the continuum-regulated ones. This can be
done either perturbatively (worrying about how convergent the calculation is)
or nonperturbatively.

The quenched approximation is a real problem. Recall that the flavor symmetry of
quenched QCD is not $SU(3)_L\otimes SU(3)_R$, it is the graded algebra 
$SU(N|N)_L\otimes SU(N|N)_R$. Consequently,
the flavor content of penguin operator  $O_6$ is altered: it
 came from running short distance physics down to
long distance using unquenched QCD evolution, but then it is going to be 
evaluated in quenched approximation. Extra low energy constants appear,
which would not be present in full QCD\cite{Golterman:2001qj}.

As a practical consequence, matrix elements of these operators have a chiral
 extrapolation similar to that of $B_K$,
\bee
< \pi |O_i | K> = C (1 + {{\xi m_{PS}^2 } \over {(4\pi f)^2}} \ln (m_{PS}^2) )
   + b m_{PS}^2 .
\ee
Golterman and Pallente\cite{Golterman:2001qj}
 computed the coefficients for chiral logarithms
in quenched and partially-quenched QCD,
and found that $\xi=0$ for  $O_8$ (in quenched approximation and in the
 degenerate-mass limit). This is different from in full QCD.
After seeing the rather large curvature of  $B_K$ in Fig. \ref{fig:combinedbk}, plausibly
 effect of a chiral logarithm,
this discovery is cause for unease.

Another  peculiarity of the quenched approximation is that
 the B-parameters for many of these operators
go to zero in the chiral limit, because $m_{PS}^2/m_q$
 diverges. Recall that the B-parameters
for these operators are defined as the ratio of the operator to its vacuum-saturated approximation.
For left-right operators,
\bee
B_i^{3/2} = {{\langle K|O_i^{3/2}|\pi\rangle} \over{
c_P \langle \pi | \bar \psi \gamma_5 \psi |0 \rangle
 \langle 0 | \bar \psi \gamma_5 \psi |K \rangle
+ c_A \langle \pi | \bar \psi\gamma_\mu \gamma_5 \psi |0 \rangle
 \langle 0 | \bar \psi \gamma_\mu\gamma_5 \psi |K \rangle }}
\ee
where the $c_i$'s are numerical coefficients.
The PCAC relation says that
 $\langle 0 | \bar \psi \gamma_5 \psi |PS \rangle ={1\over 2}
  m_{PS}^2/m_q f_{PS}$.
The divergence of $(m_{PS}^2/m_q)^2$ in the denominator is
 not compensated by any singular behavior in the
numerator.

My conclusion is that we should be happy that people are building all the
tools to do these calculations, and that one shouldn't take any of the quenched
numbers seriously.
RBC and CP-PACS are doing simulations in full QCD, and it will be interesting
to see what they find when they
revisit $\epsilon'/\epsilon$ in the next few years. I suspect that $\epsilon'/\epsilon$
will teach us more about how to do calculations than about whether the Standard Model
is right or not.

\section{Instead of conclusions}

Lattice QCD is in a period of rapid evolution. The quenched approximation is fading
out as a vehicle for precise calculations of quantities to be directly compared
with experiment.
It was helped on its way by algorithms for simulating QCD while
preserving flavor and chiral symmetry, which
cause one to attempt to do problems whose analysis forces one to confront
the fundamental differences between quenched and full QCD. Simulations are beginning to
reveal the presence of quenched chiral logarithms at small
quark mass.

Full QCD simulations are being used in phenomenology as never before.
The authors of Ref. \refcite{Davies:2003ik} are carrying out an extensive program
of light and heavy flavor physics using Asqtad dynamical fermions.
They argue that essentially all the entries of the CKM matrix (except $V_{tb}$)
can be impacted by lattice calculations whose systematics are well understood.
Asqtad fermions will eliminate the unknown quenching systematic.
Results from these simulations are already appearing, and more will appear within a year.

A big part of the lattice QCD computing resources in this country will go into
the construction of large sets of Asqtad lattices, which the idea of
making the data sets publicly available.
Other people will use their configurations as ``background fields''
for other matrix element calculations. In many cases the cost of the simulations
which are needed directly to compute a particular matrix element
is much smaller than the cost of generating the configuration
with its dynamical light quarks. Quenching systematics would presumably
be a thing of the past.
 These calculations promise
fantastic improvement in the quality of lattice determinations
of hadronic matrix elements.

The MILC collaboration is also planning to push to smaller lattice spacing and
to smaller nonstrange quark mass. Recall that their present simulations are
done at $m_{u,d}/m_s \simeq 1/5$, while in the real world the ratio about 1/20.
The smallest lattice spacing in their simulations is about 0.09 fm.
They estimate that the cost of this data set is about 0.8 Petaflop-hours.
 The cost of dropping the quark mass by a factor of two
at fixed physical volume, or of dropping the lattice spacing at fixed quark mass, 
as given by Eq. \ref{COST}, are
each less than a factor of 16 increase  over present simulations.
They believe that a combination of these two simulations will allow
extrapolations to the continuum and chiral limit at the one per cent level.

This won't happen in a year, but it is not completely unreasonable to think about
doing it. The computer resources will be there in the next decade.

However, I think that without some better understanding of the ``$\det^{1/4}$''
replacement, all these results will be clouded by uncertainty: are
these simulations fundamentally correct, or not? And I don't think that
the answer to this question will come from numerics, at least not ``direct'' numerics
in four dimensions,
like spectroscopy or matrix element calculations.
I believe it is an issue of principle.

Dynamical simulations with domain wall or overlap fermions need a lot of
development
before they can be attempted with finite computer resources.

Lattice QCD at the end of 2004 promises to be quite different from lattice QCD at the end of 2003.



\section*{Acknowledgements}

I would like to thank 
O.~B\"ar,
C.~Bernard,
C.~DeTar,
S.~D\"urr,
M.~Golterman,
S.~Gottlieb,
E.~Gregory,
A.~Hasenfratz,
P.~Hasenfratz,
U.~Heller.
K.~Jansen,
F.~Knechtli,
M.~L\"uscher,
J.~Osborne,
Y.~Shamir,
S.~Sharpe,
R.~Sugar,
D.~Toussaint,
G.~Veneziano,
and
P.~Weisz,
for their help preparing this review, or for discussions which influenced my thinking
about topics in it. I am grateful to M. Golterman
Y. Shamir for a close reading of the
manuscript.
This work was supported by the U.~S. Department of Energy.







\end{document}